\documentclass[acmtog]{acmart}

\acmSubmissionID{1111} 
\setcitestyle{nosort,square} 

\usepackage{xcolor, colortbl}
\usepackage{tikz}
\usetikzlibrary{shapes.geometric, arrows}
\usepackage{graphicx}

\copyrightyear{2026}
\acmYear{2026}
\setcopyright{cc}
\setcctype{by-nc-nd}
\acmConference[SIGGRAPH Conference Papers '26]{Special Interest Group on Computer Graphics and Interactive Techniques Conference Conference Papers}{July 19--23, 2026}{Los Angeles, CA, USA}
\acmBooktitle{Special Interest Group on Computer Graphics and Interactive Techniques Conference Conference Papers (SIGGRAPH Conference Papers '26), July 19--23, 2026, Los Angeles, CA, USA}
\acmDOI{10.1145/3799902.3811188}
\acmISBN{979-8-4007-2554-8/2026/07}

\usepackage{bm}

\usepackage{hyperref}
\usepackage{url}
\usepackage{comment} 
\usepackage{multirow}
\usepackage{subcaption}
\usepackage{dblfloatfix} 
\usepackage{tikz}
\usetikzlibrary{spy}
\newlength{\imagewidth}
\usepackage{booktabs} 
\usepackage{multirow} 
\usepackage{graphicx} 

\definecolor{orange}{rgb}{1.0,0.3,0.1}
\definecolor{blue}{rgb}{0.1,0.1,0.9}
\definecolor{purple}{rgb}{0.5,0.0,0.97}
\definecolor{cyan}{rgb}{0.0,0.6,0.8}
\definecolor{green}{rgb}{0.3,0.56,0.0}

\def\showcomments{0} 
\if\showcomments1
    \newcommand{\todo}[1]{{\color{blue} !!! TODO: #1 !!!}}
    \newcommand{\rachel}[1]{{\color{green} #1 }}
    
    \newcommand{\tnx}[1]{{\color{cyan} {#1}}}
\else
    \newcommand{\todo}[1]{{}}
    \newcommand{\rachel}[1]{#1}
    
    \newcommand{\tnx}[1]{#1}
\fi

\newcommand{\rv}[1]{{\color{black} {#1}}}

\begin{document}
\title{LagrangianSplats: Divergence-Free Transport of Gaussian Primitives for Fluid Reconstruction}

\author{Ningxiao Tao}
\email{taoningxiao@outlook.com}
\orcid{0009-0004-3616-4776}
\affiliation{
\institution{School of Intelligence Science and Technology, Peking University}
\city{Beijing}
\country{China}}

\author{Baoquan Chen}
\email{baoquan@pku.edu.cn}
\orcid{0000-0003-4702-036X}
\affiliation{
\institution{Peking University}
\city{Beijing}
\country{China}}
\authornote{corresponding authors}

\author{Mengyu Chu}
\email{mchu@pku.edu.cn}
\orcid{0000-0002-7358-433X}
\affiliation{
\institution{State Key Laboratory of General Artificial Intelligence, Peking University}
\country{China}
}\authornotemark[1]

\begin{abstract}

Reconstructing 3D fluid velocity fields from sparse 2D video observations is a highly ill-posed inverse problem, demanding both transport consistency with observed motion and physical validity under fluid laws. Existing methods typically impose these constraints through soft penalties, often leading to compromised accuracy and convergence issues. We introduce a reconstruction framework that structurally enforces both constraints. Specifically, we parameterize the reconstructed velocity using a continuous Divergence-Free Kernel representation, driving the advection of a Lagrangian 3D Gaussian Splatting representation. This formulation intrinsically guarantees both flow incompressibility and long-range transport coherence by construction. To enable the efficient optimization of such a constrained system, we introduce a novel Sliding Window scheme that propagates gradients over meaningful temporal horizons while maintaining tractable training costs. Experiments on synthetic and real-world datasets demonstrate that our method outperforms state-of-the-art baselines in both transport consistency and physical accuracy, enabling applications such as high-quality re-simulation and flow analysis. Our implementation is released at \url{https://github.com/taoningxiao/LagrangianSplats.git}.
\end{abstract}

%
%


%
%


\begin{teaserfigure}
	\centering
    \includegraphics[width=1.0\linewidth]{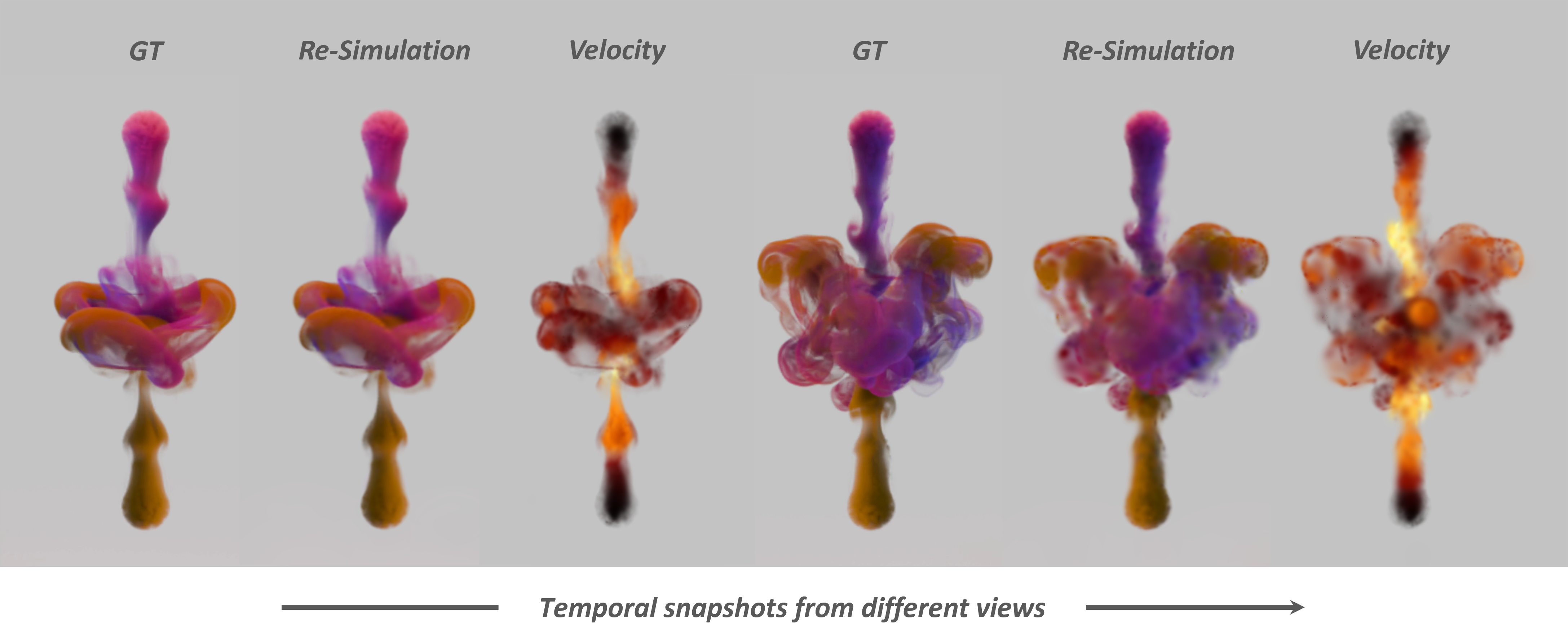}
	\vspace{-20pt}
	\caption{
Reconstruction results on the Biplume dataset, featuring highly turbulent flow dynamics. We compare our re-simulation results against the Ground Truth from two distinct viewpoints at different timestamps. Additionally, we visualize the predicted velocity magnitude using a heatmap, demonstrating our method's ability to capture intricate flow details. \rv{Quantitatively, our re-simulation achieves high fidelity, with sequence-averaged metrics of \textbf{37.70 PSNR, 0.9764 SSIM, and 0.0758 LPIPS} over the full re-simulation sequence}
\rachel{, indicating rigorous transport consistency enabled by a physically plausible velocity reconstruction that is divergence-free by construction.}
}\label{fig:teaser}
\end{teaserfigure}

\maketitle
\section{Introduction}

Fluid phenomena are ubiquitous and captivating. Reconstructing their underlying dynamics, specifically the 3D velocity field, from sparse 2D videos is a critical pursuit in computer vision and graphics. This capability unlocks broad applications ranging from \rachel{special effects~\cite{gregson2014capture} and aerodynamics analysis~\cite{rosset2023interactive} to high-fidelity forecasting~\cite{bi2023accurate}}. 
However, this task presents a formidable ill-posed inverse problem due to the inherent ambiguity of inferring 3D flow from 2D visual inputs. A successful reconstruction of the velocity field must simultaneously satisfy two rigorous criteria.
First, transport consistency requires that the reconstructed velocity field correctly advects the smoke density to reproduce the observed temporal evolution. Second, physical validity demands that the recovered flow obeys fundamental fluid laws, such as incompressibility and momentum conservation. Achieving both objectives within a unified framework remains a significant challenge.

Existing methods struggle to enforce strict transport consistency. \rachel{Eulerian approaches} like PINF~\cite{chu2022pinf} and HyFluid~\cite{yu2023inferring} model density and velocity as neural fields. 
While demonstrating the potential of joint optimization, 
\rachel{their Eulerian nature lacks explicit long-term temporal correspondence. Their short-range advection supervision often traps velocity optimization in local minima, preventing efficient convergence to globally accurate transport.

Several methods attempt to improve long-term transport consistency.}
PICT~\cite{wang2024physics} enforces consistency using neural trajectory representations. However, it relies on aggressive regularization to ensure trajectory validity, forcing the reconstruction toward overly smoothed dynamics and suppressing turbulent transport.
%
A theoretically more robust solution is GlobalTransport~\cite{franz2021global}, which defines the smoke reconstruction by explicitly advecting a canonical initial volume throughout the entire sequence. However, its reliance on differentiable rendering and advection makes it computationally prohibitive, restricting its use to scenarios with known lighting conditions and obstacle geometry. 
\rachel{As a scene representation, 3D Gaussian Splatting (3DGS)~\cite{kerbl2023gaussian_splatting} offers a promising Lagrangian perspective.}
FluidNexus~\cite{gao2025fluidnexus} adapts this paradigm for single-view smoke reconstruction by advecting Gaussians through time. \rachel{Despite its Lagrangian nature, it relies on greedy frame-to-frame velocity estimation, which discards} 
valuable long-range gradients needed to disambiguate complex flow dynamics.

In parallel to transport issues, ensuring physical validity remains a critical bottleneck. Whether employing implicit neural fields~\cite{chu2022pinf,yu2023inferring,wang2024physics} or Position-Based Fluid representations~\cite{gao2025fluidnexus}, these approaches typically supervise the velocity field using soft PINN~\cite{RAISSI2019} penalties derived from fluid laws. However, optimization based on soft constraints is often difficult to converge and cannot strictly guarantee physical plausibility.
\rachel{In particular, failure to strictly enforce the divergence-free condition introduces degeneracies: the reconstruction may explain temporal changes through artificial sources and sinks rather than through material transport. This ambiguity corrupts the velocity recovery and limits physical interpretability.}

\rachel{We present a unified reconstruction framework that enforces both constraints by construction.}
To ensure physical validity, we parameterize the velocity field using the Divergence-Free Kernel (DFK) representation~\cite{ni2025representing}. Eliminating the need for soft penalties, this approach inherently guarantees incompressibility and effectively avoids nonphysical density hallucinations.
To achieve transport consistency, we employ a Lagrangian approach where dynamic smoke is represented by 3D Gaussian primitives advected from the initial frame using the DFK velocity field. While optimizing the full dependency chain offers a strong global constraint, it is computationally prohibitive. Therefore, we introduce a Sliding Window optimization strategy.
This mechanism restricts gradient propagation to effective temporal horizons, preserving essential long-range transport logic while ensuring computational tractability.

Our main contributions are summarized as follows:
\begin{itemize}
\item We propose a strictly divergence-free fluid velocity field reconstruction framework by integrating 3DGS with the DFK representation, achieving results that are both transport-consistent and physically accurate.
\item We introduce a Sliding Window optimization mechanism that enforces Lagrangian transport consistency over meaningful temporal horizons, enabling effective long-range gradient propagation for accurate flow dynamics recovery.
\item We demonstrate state-of-the-art performance in velocity field reconstruction on both synthetic and real datasets, facilitating downstream applications such as high-fidelity re-simulation and quantitative flow analysis.
\end{itemize}
\section{Related Work}

\paragraph{Fluid Reconstruction from Sparse Views.}
Reconstructing 3D fluid dynamics is a problem of significant interest across both computer graphics and scientific research. Early scientific approaches like structured light~\cite{gu2012compressive, ji2013reconstructing, Atcheson2008OEF} and PIV~\cite{grant1997particle, elsinga2006tomographic, xiong2017rainbow} relied on specialized hardware, limiting them to laboratory settings. In graphics, efforts shifted towards reconstructing fluids from multi-view RGB videos input. ScalarFlow~\cite{eckert2019scalarflow} demonstrated this by integrating differentiable fluid solvers into the reconstruction loop. Notably, GlobalTransport~\cite{franz2021global} introduced a strict Lagrangian formulation, defining the smoke state by explicitly advecting a canonical initial volume. While this guarantees long-term transport consistency, its reliance on traditional differentiable volume rendering incurs computational bottlenecks.

Implicit Neural Representations (INRs) offer a promising continuous alternative to traditional discretizations. This paradigm has achieved remarkable success across geometry modeling~\cite{park2019deepsdf, saito2019pifu, mescheder2019occupancy,peng2020convolutional}, neural rendering~\cite{mildenhall2020nerf, barron-ICCV2021-mipnerf, muller-SIG2022-instantngp}, and physics simulation~\cite{chen2023implicit}. In the context of fluid reconstruction, recent works have leveraged INRs to jointly model density and velocity fields. PINF~\cite{chu2022pinf} and HyFluid~\cite{yu2023inferring} incorporate physics-informed losses to enhance physical plausibility, while NeuSmoke~\cite{qiu2024neusmoke} combines 3D neural fields with 2D refinement for efficiency. To enforce temporal coherence, PICT~\cite{wang2024physics} adopts a trajectory-based representation for long-term supervision. However, its over-constrained formulation often limits its ability to capture fine-grained turbulent details. More broadly, INRs generally depend on PINN-based objectives~\cite{RAISSI2019}, whose ill-conditioned optimization landscape often hinders effective training and limits reconstruction fidelity~\cite{rathore2024challenges,cao2025analysis}.

\paragraph{3D Gaussian Splatting for Dynamic and Fluid Scenes.}
The advent of 3D Gaussian Splatting~\cite{kerbl2023gaussian_splatting} has catalyzed a shift toward dynamic scene reconstruction. Numerous approaches~\cite{luiten2024dynamic,yang2024deformable,wu20244d,duan20244d,hong2025physics} extend 3DGS to the temporal domain, primarily prioritizing visual fidelity in general scenes.

To enable physically plausible fluid reconstruction, recent works have integrated physics-informed supervision into the 3DGS framework. GaussFluids~\cite{du2025gaussfluids} incorporates a density-based soft constraint to guide particle distribution and encourage incompressibility, yet it lacks explicit momentum supervision governed by the Navier-Stokes equations. FluidGS~\cite{xie2025fluidgs} transfers Gaussian density fields to Eulerian grids to apply PINN constraints; however, this supervision remains a soft regularization restricted to a localized temporal horizon. 
Most closely related to our work is FluidNexus~\cite{gao2025fluidnexus}, which sharing a similar Lagrangian motivation to advect 3DGS using Position-Based Fluid representation~\cite{macklin2013position}. However, FluidNexus adopts a sequential, frame-by-frame estimation strategy, which discards valuable long-range gradients from future frames. Our approach diverges by addressing the problem through a global transport perspective, capturing long-range temporal dependencies often overlooked by greedy, sequential updates.

\paragraph{Divergence-Free Velocity Representation.}
Divergence-free motion is critical for realistic fluid simulation. Traditional solvers typically enforce incompressibility via pressure projection~\cite{zhu2005animating, jiang2015affine,bridson2015fluid,stam2023stable}. Recent research has explored representations that are divergence-free by construction, such as maintaining vector potentials~\cite{chang2021curl, lyu2024wavelet} or employing specialized grid interpolation schemes~\cite{nabizadeh2024fluid, roy2024higher}. In the data-driven domain, efforts have been made to encode these physical constraints into neural architectures. For instance, \citet{kim2019deep} trained Convolutional Neural Networks to predict velocity potentials, while \citet{richter2022neural} proposed neural parameterizations that inherently satisfy the divergence-free condition. Unlike these grid-based or implicit approaches, we adopt the Divergence-Free Kernel~\cite{ni2025representing}. Its mesh-free, analytically incompressible nature naturally complements the Lagrangian 3DGS, ensuring continuous fields suitable for sparse particle advection.


\section{Background} \label{sec:background}

In this section, we briefly review the foundational representations that underpin our framework: 3D Gaussian Splatting (Sec.~\ref{sec:bg_3dgs}) and Divergence-Free Kernels (Sec.~\ref{sec:bg_dfk}).

\subsection{3D Gaussian Splatting}
\label{sec:bg_3dgs}
3D Gaussian Splatting~\cite{kerbl2023gaussian_splatting} represents a scene as a set of 3D anisotropic Gaussians. Each Gaussian $k$ is parameterized by its mean position $\boldsymbol{\mu}_k \in \mathbb{R}^3$, covariance matrix $\boldsymbol{\Sigma}_k$, opacity $\alpha_k \in [0, 1]$, and view-dependent color represented by spherical harmonics. To ensure positive semi-definiteness, the covariance matrix $\boldsymbol{\Sigma}_k$ is further decomposed into a scaling matrix $\mathbf{S}_k$ and a rotation matrix $\mathbf{R}_k$, such that $\boldsymbol{\Sigma}_k = \mathbf{R}_k \mathbf{S}_k \mathbf{S}_k^\top \mathbf{R}_k^\top$.

To render an image from a specific viewpoint, the 3D Gaussians are projected onto the 2D image plane. The pixel color $C(\mathbf{p})$ is computed using volumetric $\alpha$-blending, sorting the Gaussians from front to back:
\begin{equation}
    C(\mathbf{p}) = \sum_{i \in \mathcal{N}} c_i \sigma_i \prod_{j=1}^{i-1} (1 - \sigma_j),
\end{equation}
where $c_i$ is the color of the $i$-th Gaussian, and $\sigma_i$ is the effective opacity computed by multiplying the learned opacity $\alpha_i$ with the projected 2D Gaussian's probability density. This explicit representation allows for real-time rasterization and efficient differentiability, which is crucial for our long-range transport optimization.

\subsection{Divergence-Free Kernel}
\label{sec:bg_dfk}

DFK~\cite{ni2025representing} represent velocity field $\mathbf{v}(\mathbf{x})$ over a spatial domain as a linear combination of matrix-valued kernels $\boldsymbol{\psi}_i$:
\begin{equation}
    \mathbf{v}(\mathbf{x}) = \sum_{i} \boldsymbol{\psi}_i(\mathbf{x}) \, \boldsymbol{\omega}_i,
\label{eq:dfk}
\end{equation}
where $\boldsymbol{\omega}_i \in \mathbb{R}^3$ is the trainable weight vector associated with the $i$-th kernel centered at $\mathbf{x}_i$. The matrix-valued kernel $\boldsymbol{\psi}_i: \mathbb{R}^3 \to \mathbb{R}^{3 \times 3}$ is derived from a scalar radial basis function via a second-order differential operator:
\begin{equation}
    \boldsymbol{\psi}_i(\mathbf{x}) = (-\mathbf{I} \nabla \cdot \nabla + \nabla \nabla^\top) \, \phi_i(\mathbf{x}),
\end{equation}
where $\mathbf{I}$ is the identity matrix, $\nabla \nabla^\top$ denotes the Hessian operator, and $\phi_i(\mathbf{x})$ is a scalar kernel defined as:
\begin{equation}
    \phi_i(\mathbf{x}) = \phi\left(\frac{\|\mathbf{x} - \mathbf{x}_i\|}{h_i}\right),
\end{equation}
with $h_i$ being the support radius. Following~\citet{ni2025representing}, we employ the Wendland $C^4$-continuous piecewise-polynomial function~\cite{wendland1995piecewise} as the radial basis function $\phi(r)$.

By construction, the divergence of the resulting field $\mathbf{v}(\mathbf{x})$ in Eq.~\eqref{eq:dfk} is guaranteed to be identically zero for any choice of weights $\boldsymbol{\omega}_i$. Beyond ensuring incompressibility, this representation offers several desirable properties, including compact support, positive definiteness, and second-order differentiability, making it an ideal parameterization for physically plausible fluid reconstruction.

\section{Method}

\begin{figure*}[ht]
    \centering
    \includegraphics[width=.9\linewidth,trim=0cm 0.0cm 0.0cm 0.0cm,clip]{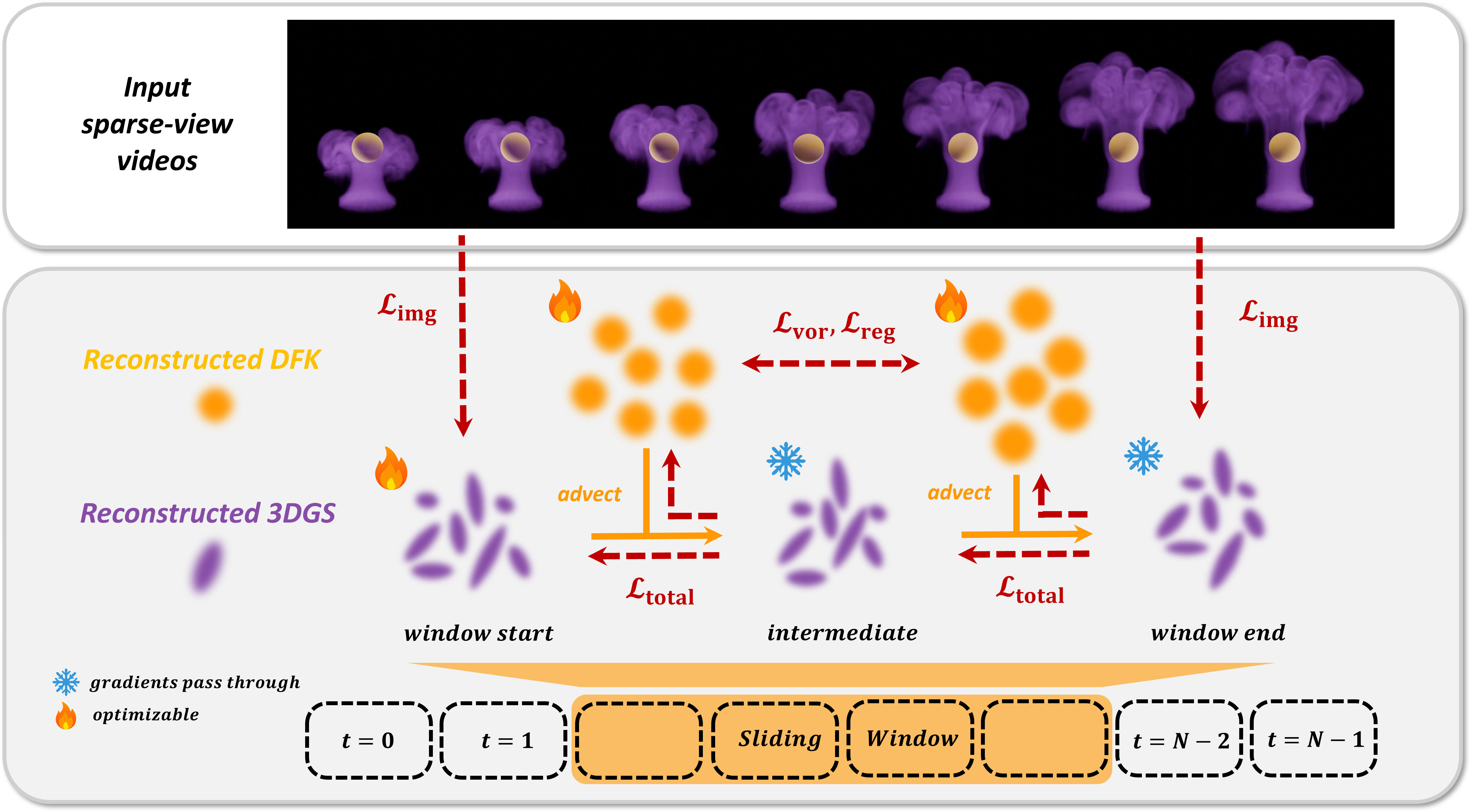}
    \vspace{-6pt}
    \caption{\textbf{Overview of our proposed framework.} We reconstruct smoke dynamics from sparse-view videos using a hybrid representation of 3DGS (purple splats) for density and DFK (orange nodes) for velocity. To ensure transport consistency, we employ a Sliding Window strategy. Within the active window, the smoke state at intermediate frames is explicitly defined by advecting the Gaussians from the window's start. As indicated by the icons, the Gaussian parameters at the window start and the velocity fields are optimizable, while the advected states are derived but allow gradients to pass through. Consequently, gradients from the image loss $\mathcal{L}_{\text{img}}$ back-propagate through the advection chain (red dashed arrows), jointly refining the initial Gaussian attributes and the velocity field with long-term temporal foresight. Physical validity is further enhanced via vorticity and regularization losses $\mathcal{L}_{\text{vor}}$ and $\mathcal{L}_{\text{reg}}$.}
    \label{fig:pipeline}
\end{figure*}

This section describes our reconstruction framework. Given sparse multi-view videos of smoke, our goal is to jointly recover both the evolving smoke and the driving velocity field. As illustrated in Fig.~\ref{fig:pipeline}, our physics-integrated framework seamlessly combines the efficient differentiable rendering of 3DGS with the physical guarantees of DFK-parameterized velocity. At its core is a Lagrangian-based optimization scheme that drives advection via a strictly divergence-free velocity field, inherently ensuring both transport consistency and physical plausibility by construction.

We first define our hybrid representation in Sec.~\ref{sec:advected_gaussians}, coupling 3DGS geometry with the DFK velocity parameterization. Building on this model, Sec.~\ref{sec:sliding_window} introduces our Sliding Window strategy, an algorithm designed to capture long-range dynamics effectively by propagating gradients over manageable temporal horizons. Finally, Sec.~\ref{sec:optimize_obj} outlines the optimization objectives, comprising reconstruction terms and physics-informed regularizers to guide the joint recovery process.

\subsection{Modeling Smoke with Advected Gaussians}
\label{sec:advected_gaussians}

We represent the evolving smoke at time $t$ as a set of 3D Gaussians,
\begin{equation}
    \mathcal{G}_t = \{ \mathbf{G}_i^t \}_{i=1}^{N},
\end{equation}
\rachel{Crucially, rather than optimizing their positions as free variables, we treat them as passive tracers carried by the DFK-parameterized velocity field}$\mathbf{v}(\mathbf{x}, t)$.
Specifically, the position of each Gaussian at time $t$ is determined by advecting its initial state from the first frame:
\begin{equation}
    \frac{d \boldsymbol{\mu}_i}{dt} = \mathbf{v}(\boldsymbol{\mu}_i, t), 
    \qquad
    \boldsymbol{\mu}_i^t = \mathcal{A}_{0 \rightarrow t}(\boldsymbol{\mu}_i^0, \mathbf{v}),
    \label{eq:adv}
\end{equation}
where $\mathcal{A}_{0 \rightarrow t}(\cdot)$ denotes the Lagrangian advection operator. In this way, the movement of Gaussians is explicitly constrained by the advection process, providing a physically grounded and differentiable link between the reconstructed smoke and the underlying flow dynamics.

\subsection{Sliding Window Optimization}
\label{sec:sliding_window}

We propose a sliding window strategy to jointly optimize the smoke and the underlying velocity field from sparse video inputs. Given a sequence of $N$ frames and a window size $w$, the optimization proceeds in two phases: an initial warm-up phase and a subsequent sliding phase.

\paragraph{Initial Warm-up Phase.}
\rv{
We begin by reconstructing the smoke at the first frame ($t=0$) using standard 3D Gaussian Splatting to provide an initial guess. The optimization then focuses on the initial window spanning frames $[0, w]$, where the first frame ($t=0$) acts as the \textit{anchor frame}. Within this window, the intrinsic attributes (e.g., opacity) of the anchor frame are actively optimized as trainable parameters to refine the initial guess. For any subsequent frame $t \in (0, w]$, the positions of Gaussians are derived via the advection operator $\mathcal{A}_{0\rightarrow t}$ defined in Eq.~\eqref{eq:adv}. To enforce short-term physical consistency, the intrinsic attributes of these advected Gaussians strictly inherit the updated values from the anchor and remain constant along the intra-window trajectory.}

To ensure training stability, we adopt a progressive strategy. Starting from a short horizon, we gradually extend the advection range from $t=1$ to $t=w$. At each step, the advected Gaussians are rendered and supervised by an image reconstruction loss $\mathcal{L}_{\text{img}}$, which combines L1 and D-SSIM terms~\cite{kerbl2023gaussian_splatting}. Gradients from this loss are back-propagated analytically to update the time-varying velocity field across the window and the parameters of the Gaussians at the initial frame. 

Crucially, when aggregating gradients over the window, we account for the diminishing influence of distant physical interactions by applying a temporal discount factor $\gamma\in (0,1)$. Specifically, to update the parameters at frame $t$ using future supervision, the objective is formulated as a weighted sum:
\begin{equation}
\label{eq:discounted_loss}
\mathcal{L}_{\text{total}} = \sum_{j} \gamma^j \mathcal{L}_{\text{img}}(t+j),
\end{equation}
where $j$ represents the temporal distance from the current frame $t$ to the supervised frame $t+j$. In our implementation, we set $\gamma = 0.9$ as the default value. This discounting strategy effectively suppresses noise arising from long-range dependencies, thereby stabilizing the joint optimization of the smoke and the underlying velocity field.
\paragraph{Subsequent Sliding Phase.}
After the warm-up, we transition to the sliding phase. Consider the window shifting from the span $[s-1,s-1+w]$ to $[s,s+w]$, where the start frame index $s$ progresses from $1$ to $N-1-w $. At this stage, the Gaussians at frame $s-1$ have been optimized by the previous windows sufficiently. Since the temporal discount factor $\gamma$ renders the impact of gradients from future frames negligible on this distant past state, we fix its parameters to truncate the computational graph.

In the new window, optimization follows a similar advection logic as the initial warm-up phase, but with a key difference in handling the anchor frame $s$. To strictly enforce global Lagrangian consistency, the positions of Gaussians at frame $s$ are not optimized directly. Instead, they are constrained to be the result of advecting fixed positions from frame $s-1$ using the velocity field $\mathbf{v}_{s-1}$. Consequently, position-related gradients back-propagated to frame $s$ will flow further back, optimizing velocity $\mathbf{v}_{s-1}$. This ensures the advection trajectory connecting the windows remains physically continuous.

\rv{
While positions are strictly constrained by the advection from the previous window, the intrinsic attributes (e.g., opacity) of the new anchor frame $s$ are actively optimized. We initialize these attributes using values from frame $s-1$ to promote temporal smoothness, but allow them to be optimized as free variables. This enables the model to capture progressive appearance changes like the diffusion of the smoke. Crucially, the remaining frames in the window ($t \in (s, s+w]$) are advected from this new anchor, strictly inheriting its updated intrinsic attributes just as in the warm-up phase. Finally, as the majority of frames in the new window are already pre-optimized, we bypass the progressive schedule and directly optimize over the full window width $w$. This process repeats until the end of the sequence, efficiently scaling our algorithm to arbitrary video lengths while enforcing a continuous physical trajectory.
}

Additionally, to model continuous smoke emission, we designate a specific inflow region. When a new frame first enters the sliding window, we augment the model by injecting new Gaussians into this region. Once added, these new Gaussians are treated identically to the other advected ones, fully participating in the standard velocity-driven optimization process.

\subsection{Optimization Objectives}
\label{sec:optimize_obj}

The overall optimization objective is designed to jointly ensure high-fidelity visual reconstruction and physical plausibility. It comprises two primary components: reconstruction losses that supervise the smoke's geometry and appearance against observed videos, and physics-informed velocity constraints that ensure the underlying velocity field to adhere to fluid dynamics principles.

\paragraph{Reconstruction Losses.}
As defined in Eq.~\eqref{eq:discounted_loss}, our primary supervision comes from the discounted image reconstruction loss. Given a rendered frame and ground truth, the per-frame loss is a combination of L1 distance and D-SSIM~\cite{kerbl2023gaussian_splatting}:
\begin{equation}
\mathcal{L}_{\text{img}} = (1 - \lambda_{\text{ssim}}) \mathcal{L}_1 + \lambda_{\text{ssim}} \mathcal{L}_{\text{D-SSIM}},
\end{equation}
where we set $\lambda_{\text{ssim}} = 0.2$.

Additionally, to prevent visual artifacts caused by overly skinny Gaussian particles, we also apply the anisotropic regularization term $\mathcal{L}_{\text{aniso}}$ proposed by ~\citet{xie2025fluidgs}:
\begin{equation}
    \mathcal{L}_{\text{aniso}} = \sum (||s_x - s_y||^2 + ||s_y - s_z||^2 + ||s_x - s_z||^2),
\end{equation}
where $s_x, s_y, s_z$ are Gaussian particles' scaling factor in each axes.

\paragraph{Physics-Informed Velocity Constraints.}
While the DFK representation guarantees the velocity field is divergence-free, we further enforce physical plausibility through two loss terms.

First, to penalize spurious motion in empty regions where no smoke density exists to constrain the flow, we impose a regularization based on the $L_1$ norm of the velocity magnitude:
\begin{equation}
\mathcal{L}_{\text{reg}} = \sum_x|\mathbf{v(x)}|_1.
\end{equation}
This term encourages the velocity field to remain zero in the absence of smoke, guiding the optimization toward a minimal-energy solution.

Second, to capture realistic fluid dynamics, we enforce the momentum conservation described by the Navier-Stokes equations. Since our field is analytically incompressible, we can work directly with the Vorticity Transport Equation~\cite{cottet2001vortex}. We minimize the residual of the vorticity evolution:
\begin{equation}
\mathcal{L}_{\text{vor}} = \left| \frac{\partial \boldsymbol{\omega}}{\partial t} + (\mathbf{v} \cdot \nabla)\boldsymbol{\omega} - (\boldsymbol{\omega} \cdot \nabla)\mathbf{v} \right|_1,
\end{equation}
where $\boldsymbol{\omega} = \nabla\times \mathbf{v}$ is the vorticity. By combining the hard divergence-free constraint of DFK with this soft vorticity transport constraint, our method effectively reconstructs a velocity field that adheres to the full incompressible Navier-Stokes dynamics.

\paragraph{Total Objective.}
Finally, we integrate the aforementioned terms into a unified objective function to guide the joint optimization process. The total loss is formulated as a weighted sum:
\begin{equation}
\mathcal{L} = \mathcal{L}_{\text{total}} + \lambda_{\text{aniso}} \mathcal{L}_{\text{aniso}} + \lambda_{\text{reg}} \mathcal{L}_{\text{reg}} + \lambda_{\text{vor}} \mathcal{L}_{\text{vor}},
\end{equation}
where $\mathcal{L}_{\text{total}}$ is the temporally discounted objective from Eq.~\eqref{eq:discounted_loss}. The hyperparameters $\lambda_{\text{aniso}}$, $\lambda_{\text{reg}}$ and $\lambda_{\text{vor}}$ control the relative importance of the shape regularization, kinetic energy sparsity, and vorticity transport consistency, respectively. By minimizing this composite objective, our method effectively balances the trade-off between transport consistency and physical accuracy.
\section{Experiments}

\begin{table*}[htbp]
\caption{Quantitative comparison on the ScalarSyn (synthetic) and ScalarReal (real-world) benchmarks. We evaluate transport consistency using re-simulation metrics (PSNR, SSIM, LPIPS) and assess physical validity via divergence (Div), velocity error (MSE-$v$), and directional consistency (Cos-$v$). Note that both MSE-$v$ and Cos-$v$ are computed only within the smoke volume, masked by regions where GT density is greater than $0$, to focus on the relevant dynamics. Additionally, while our method guarantees zero divergence analytically, we report the numerical divergence computed via finite differences on the grid for fair comparison with baselines. Our method consistently outperforms all baselines, achieving the highest visual fidelity and the lowest physical errors.}
\vspace{-0.5em}
\centering
\small
\begin{tabular}{lcccccccccccc}
\toprule
\multirow{2}{*}{Model} & \multicolumn{6}{c}{ScalarSyn} & \multicolumn{6}{c}{ScalarReal} \\
\cmidrule(lr){2-7} \cmidrule(lr){8-13} 
 & PSNR$\uparrow$ & SSIM$\uparrow$ & LPIPS$\downarrow$ & Div$\downarrow$ & MSE-$v$$\downarrow$ & Cos-$v$$\uparrow$ & PSNR$\uparrow$ & SSIM$\uparrow$ & LPIPS$\downarrow$ & Div$\downarrow$ & MSE-$v$$\downarrow$ & Cos-$v$$\uparrow$ \\
\midrule
PINF    & 30.06 & 0.9225 & 0.08284 & 0.004024 & 0.1465 & 0.03890 & 31.24 & 0.9635 & 0.1020 & 0.003482 & -- & -- \\
PICT    & 31.00 & 0.8998 & 0.1000 & 0.002380 & 0.1166 & 0.06472 & 31.97 & 0.9490 & 0.09932 & 0.002177 & -- & -- \\
HyFluid & 32.41 & 0.9340 & 0.08536 & 0.3268   & 0.4033 & 0.1367 & 32.62 & 0.9555 & 0.08186 & 0.003058 & -- & -- \\
FluidNexus & 31.93 & 0.9501 & 0.1087 & 0.01774   & 0.1430 & 0.05362 & 35.83 & 0.9781 & 0.07847 & 0.04148 & -- & -- \\
Ours    & \textbf{46.50} & \textbf{0.9930} & \textbf{0.02581} & \textbf{0.000099} & \textbf{0.1116} & \textbf{0.2731} & \textbf{40.00} & \textbf{0.9782} & \textbf{0.06440} & \textbf{0.000008} & -- & -- \\
\bottomrule
\end{tabular}

\label{tbl:std_bench}
\end{table*}

We evaluate our method against state-of-the-art approaches on a diverse set of synthetic and real-world scenarios. Experiments demonstrate that our framework outperforms existing methods in terms of both transport consistency and physical validity. In the following, we will first detail our experimental setup.

\paragraph{Datasets.} 
We conduct quantitative benchmarking on the ScalarSyn dataset~\cite{eckert2019scalarflow}, which provides ground-truth (GT) velocity fields, and the ScalarReal dataset~\cite{eckert2019scalarflow}, a real-world capture. \rv{Additionally, to assess robustness in challenging scenarios, we introduce three new datasets: Suzanne for smoke interacting with complex obstacle geometries over long sequences, Sphere for boundary interactions and Biplume for high turbulence.}

\begin{figure*}[htbp]
    \centering
    
    \setlength{\imagewidth}{0.166\textwidth}
    
    \def\synHeight{130pt} 
    
    \newcommand{\formattedgraphics}[3]{%
        \begin{tikzpicture}[spy using outlines={rectangle, magnification=2, connect spies}]
          \fill[black] (0, 10pt) rectangle (\imagewidth, \synHeight);
          
          \clip (0, 10pt) rectangle (\imagewidth, \synHeight);
          
          \node[anchor=south west, inner sep=0] at (0,0){\includegraphics[width=\imagewidth]{#1}};
          
          \node[anchor=north west,text=white] at (.02\imagewidth, \synHeight - 5pt){\sffamily\footnotesize\textbf{#2}};
          
          \node[anchor=north west,text=white] at (.02\imagewidth, \synHeight - 15pt){\sffamily\scriptsize #3};
        \end{tikzpicture}%
    }

    \formattedgraphics{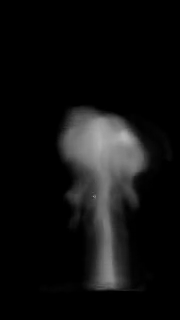}{PINF}{}%
    \hspace{-0.18cm}%
    \formattedgraphics{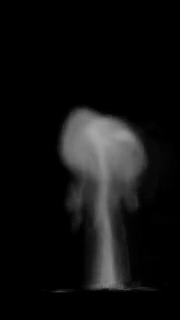}{PICT}{}%
    \hspace{-0.18cm}%
    \formattedgraphics{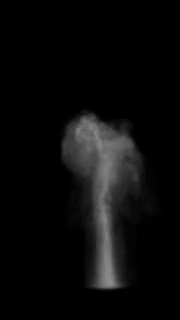}{HyFluid}{}%
    \hspace{-0.18cm}%
    \formattedgraphics{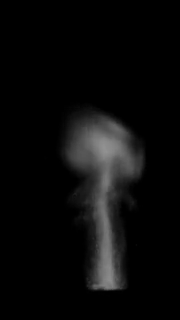}{FluidNexus}{}%
    \hspace{-0.18cm}%
    \formattedgraphics{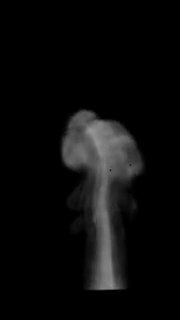}{Ours}{}%
    \hspace{-0.18cm}%
    \formattedgraphics{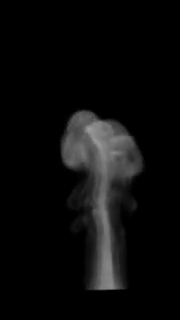}{Ground Truth}{}%
    \\
    \vspace{-6pt}
    \caption{Comparison of re-simulation results on the ScalarSyn dataset. \rv{We visualize a late-time snapshot of the smoke density advected by the reconstructed velocity fields of different methods}. Baseline methods suffer from severe density drift and dissipation, failing to maintain the smoke's shape over long durations. Our method preserves sharp structural details and correct advection trajectories, achieving superior visual fidelity that is visually indistinguishable from the Ground Truth.}
    \label{fig:resimSyn}
    
    \vspace{1em}
    
    \setlength{\imagewidth}{0.166\textwidth}
    
    \def\realHeight{140pt}
    
    \newcommand{\formattedgraphicsaa}[3]{%
        \begin{tikzpicture}[spy using outlines={rectangle, magnification=2, connect spies}]
        
        \clip (0, 0pt) rectangle (\imagewidth, \realHeight);
        
          \node[anchor=south west, inner sep=0] at (0,0){\includegraphics[width=\imagewidth,trim={2cm 2cm 1cm 12cm},clip]{#1}};
          
          \node[anchor=north west,text=white] at (.02\imagewidth, \realHeight - 15pt) {\sffamily\footnotesize\textbf{#2}};
          
          \node[anchor=north west,text=white] at (.02\imagewidth, \realHeight - 25pt) {\sffamily\scriptsize #3};
        \end{tikzpicture}%
    }  
    
    \formattedgraphicsaa{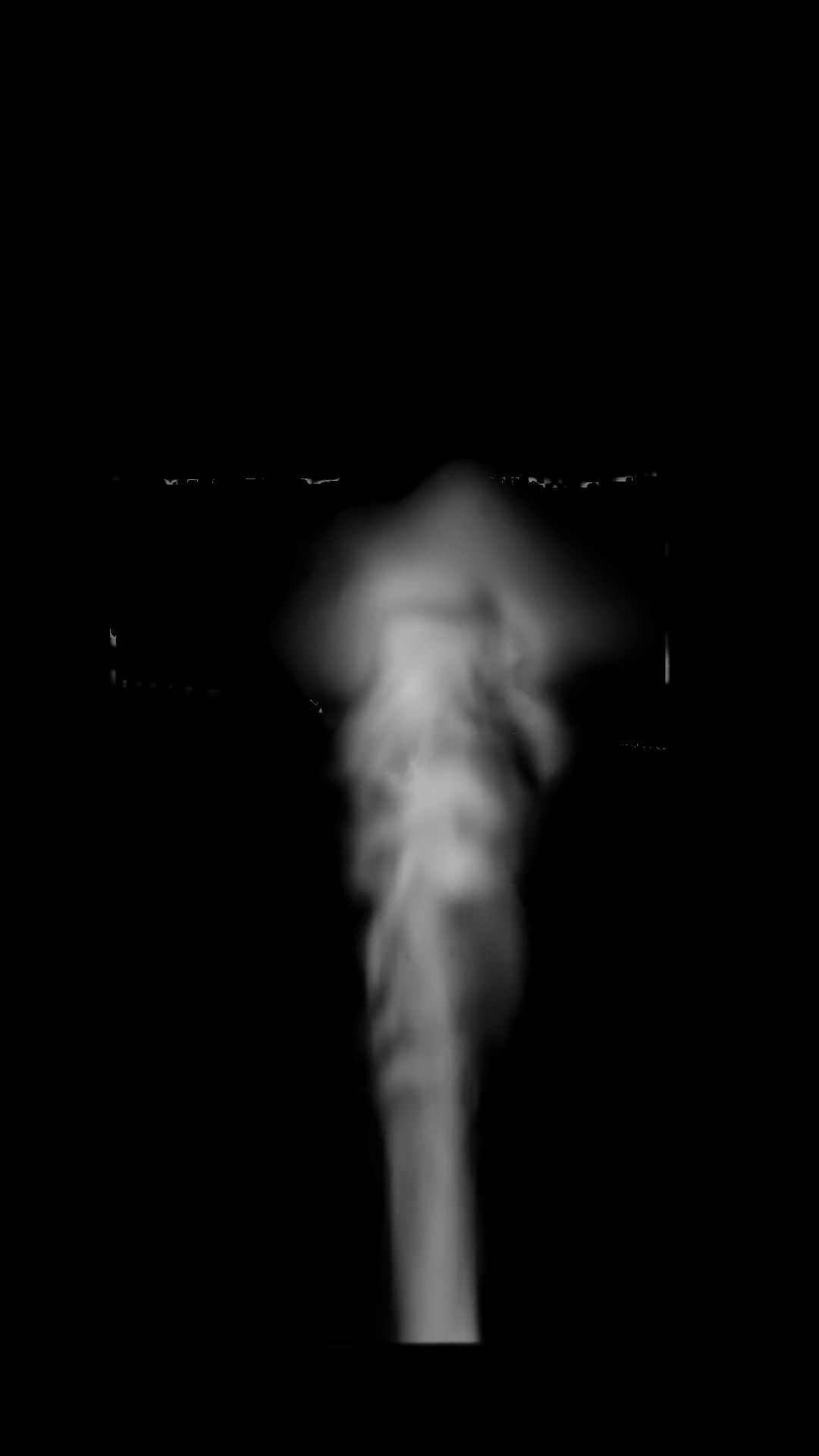}{PINF}{}%
    \hspace{-0.18cm}%
    \formattedgraphicsaa{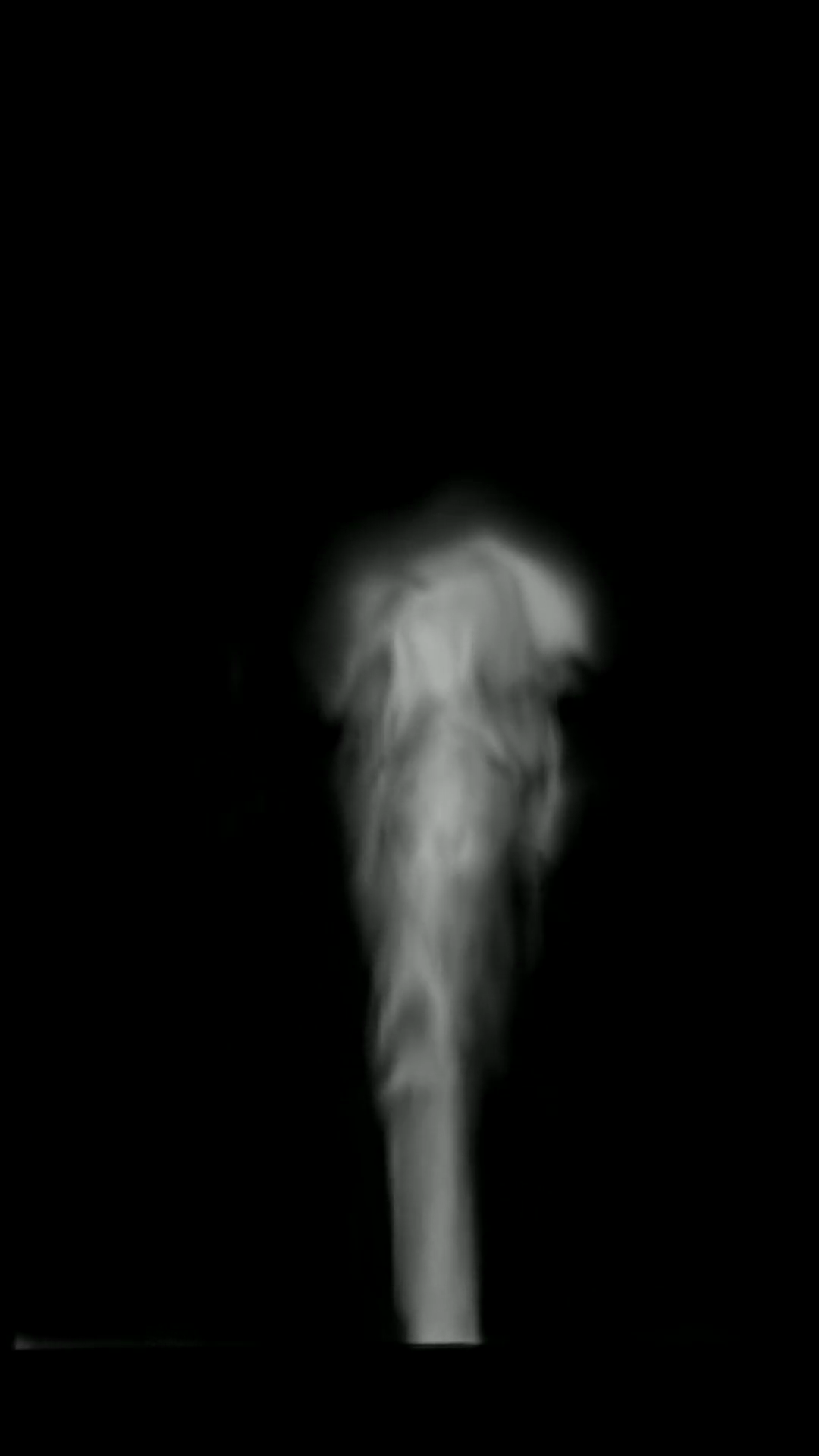}{PICT}{}%
    \hspace{-0.18cm}%
    \formattedgraphicsaa{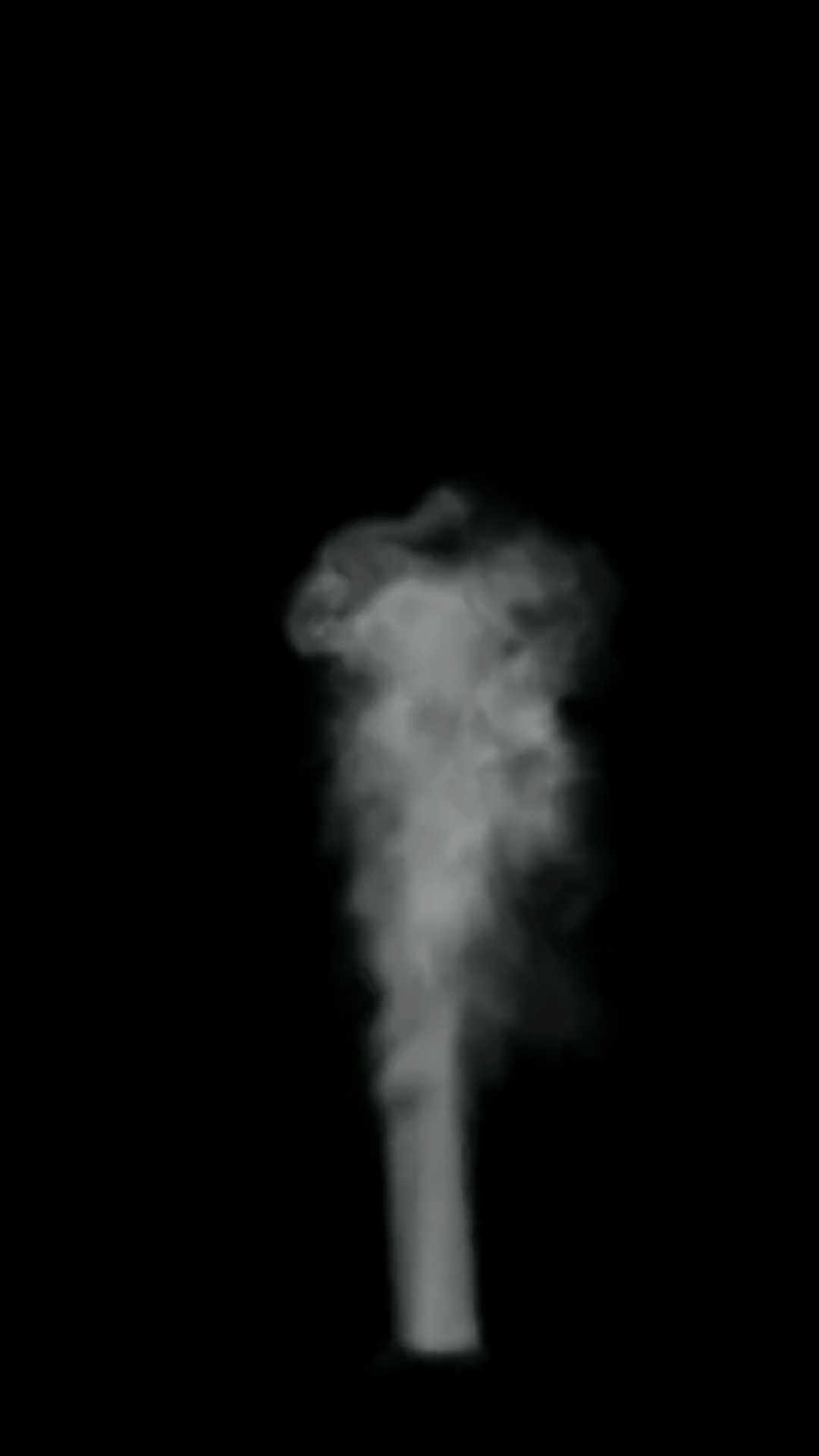}{HyFluid}{}%
    \hspace{-0.18cm}%
    \formattedgraphicsaa{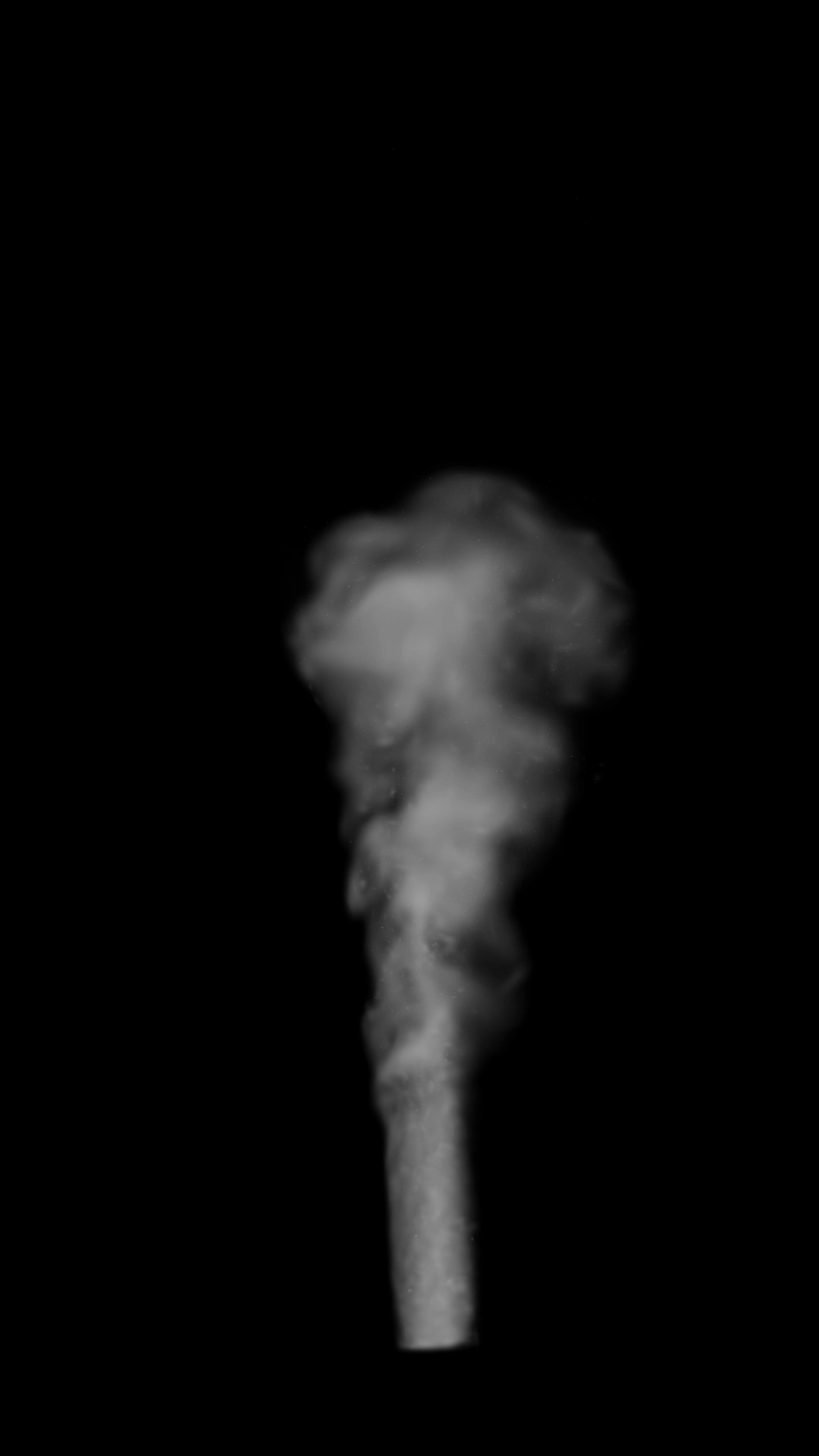}{FluidNexus}{}%
    \hspace{-0.18cm}%
    \formattedgraphicsaa{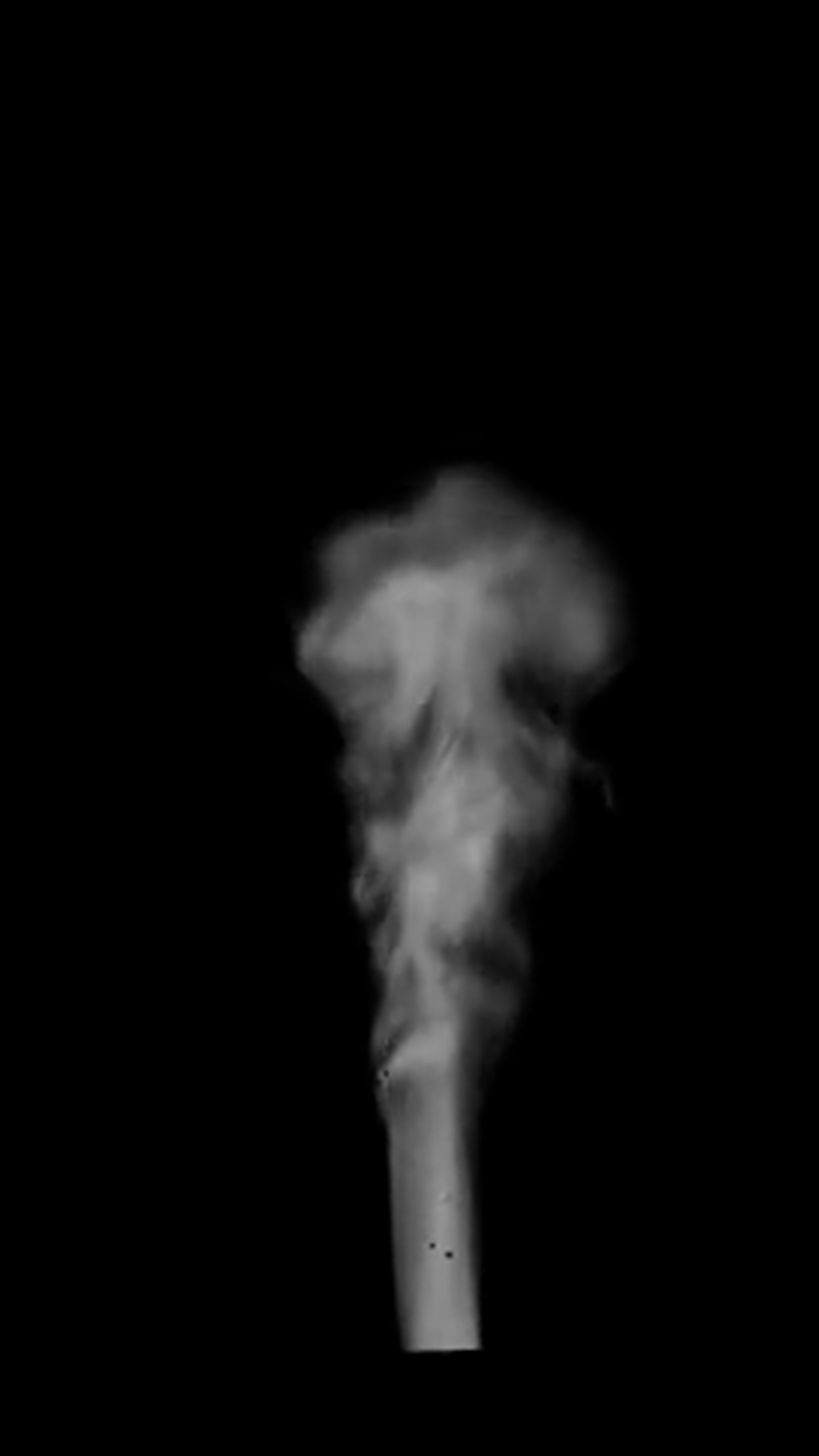}{Ours}{}%
    \hspace{-0.18cm}%
    \formattedgraphicsaa{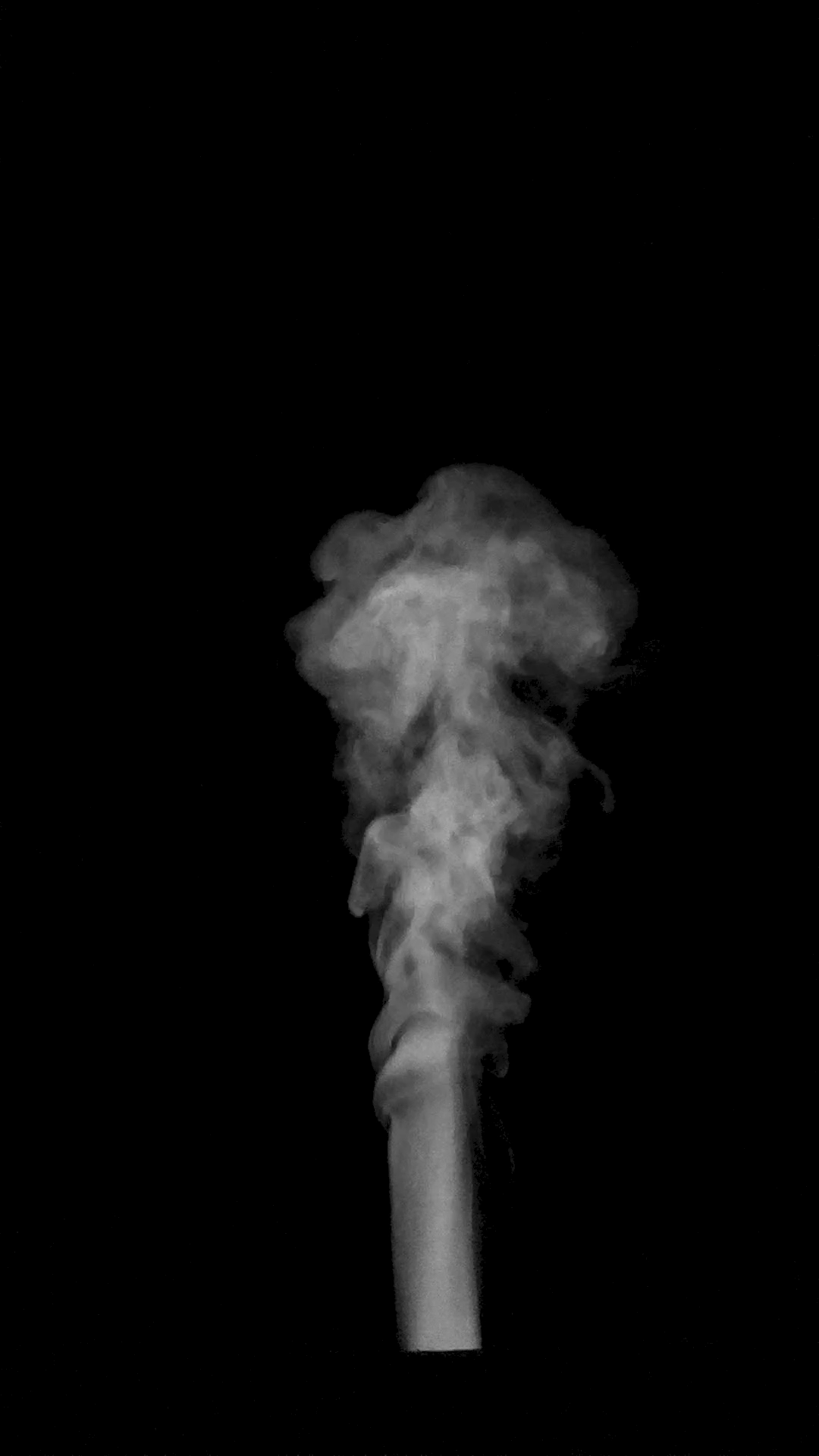}{Ground Truth}{}%
    \\
    \vspace{-0.5em}
    \captionof{figure}{Comparison of re-simulation results on the ScalarReal dataset. Despite the complexity of real-world capture, our method maintains long-term transport consistency, preserving the plume's structure significantly better than baselines. This confirms the efficacy of our Sliding Window strategy in capturing accurate Lagrangian dynamics from sparse real-world observations.}
    \label{fig:resimReal}
    \vspace{1em}
\end{figure*}

\begin{figure*}[bp]
    \centering
    \begin{minipage}[t]{\linewidth}
        \centering
        \begin{minipage}[t]{0.32\linewidth}
            \includegraphics[width=\linewidth]{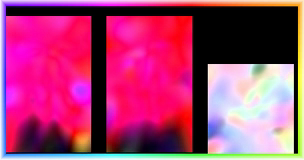}
            \vspace{-1.9em}
            \caption*{PINF}
            \includegraphics[width=\linewidth]{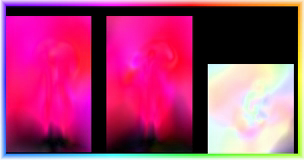}
            \vspace{-1.9em}
            \caption*{PICT}

        \end{minipage}%
        \hspace{-0.3em}
        \begin{minipage}[t]{0.32\textwidth}
            \centering
            \includegraphics[width=\textwidth]{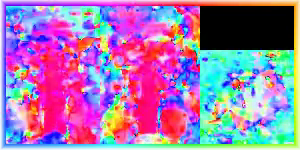}
            \vspace{-1.9em}
            \caption*{HyFluid}
            
            \includegraphics[width=\textwidth]{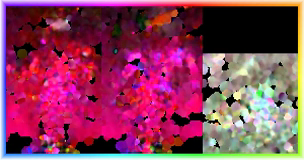}
            \vspace{-1.9em}
            \caption*{FluidNexus}
        \end{minipage}
        \hspace{-0.5em}
        \begin{minipage}[t]{0.32\textwidth}
            \includegraphics[width=\linewidth]{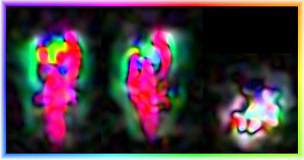}
            \vspace{-1.9em}
            \caption*{Ours}
            \includegraphics[width=\textwidth]{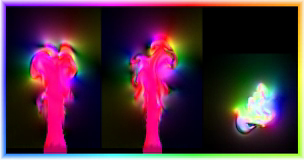}
            \vspace{-1.9em}
            \caption*{GT}
        \end{minipage}
    \end{minipage}
    \vspace{-6pt}
    \caption{Visualization of reconstructed velocity fields on the ScalarSyn dataset. It can be seen that both PINF and PICT produce overly smoothed fields, while HyFluid and FluidNexus exhibit significant noise or discontinuities. In contrast, our method reconstructs clean, coherent flow structures that closely align with the Ground Truth.}
    \label{fig:velcmpSyn}
\end{figure*}

\paragraph{Baselines.}
We compare our framework with leading fluid reconstruction methods, focusing on representative implicit neural representations: PINF~\cite{chu2022pinf}, HyFluid~\cite{yu2023inferring}, and PICT~\cite{wang2024physics}. We also include the recent 3DGS-based approach, FluidNexus~\cite{gao2025fluidnexus}, for a comprehensive comparison.

\paragraph{Evaluation Metrics.}
We assess reconstruction quality from two complementary perspectives. 

First, for transport consistency, we perform the re-simulation task proposed by~\citet{yu2023inferring}. Specifically, we advect the initial reconstructed smoke solely using the recovered velocity field over the entire sequence. 
\rv{
The re-simulated results are then rendered and compared with the corresponding ground-truth images. Given the highly sparse input views in our standard setting, we primarily evaluate these re-simulation metrics under the training views. However, to further assess our method's performance from unobserved angles, we additionally include test-view re-simulation evaluations in the Suzanne scene. For both settings, we report sequence-averaged PSNR, SSIM~\cite{wang2004image}, and LPIPS~\cite{zhang2018unreasonable}, obtained by first computing the metrics for each frame and then averaging them over the full re-simulation sequence.
}

Second, for physical validity, we measure the average divergence (Div) of the reconstructed velocity field across all datasets to quantify violation of incompressibility. \tnx{For synthetic datasets where GT velocity is available, we additionally evaluate the flow accuracy using two metrics: the Mean Squared Error (MSE-$v$) to measure value deviation, and the average Cosine Similarity (Cos-$v$) to assess directional alignment between the reconstructed and ground-truth velocities.} Both metrics are computed within the smoke volume, masked by regions where GT density is greater than zero.

\paragraph{Visualization.}
\tnx{For qualitative analysis, we visualize velocity fields via middle slices of the $xy$, $zy$, and $xz$ planes arranged side-by-side, following standard protocols~\cite{chu2022pinf,wang2024physics}. In this visualization, velocity direction is encoded as color hue and magnitude as brightness. The image border acts as a circular legend to identify directions; for instance, colors appearing at the top of the border (purplish-red) denote upward motion, whereas colors at the bottom (green) denote downward motion.}

\medskip
\rv{With the experimental setup established, the remainder of this section is organized as follows: Sec.~\ref{sec:comp_standard} presents a comprehensive comparison against baselines on standard benchmarks. To demonstrate the scalability and generalization of our approach, Sec.~\ref{sec:supp_complex_scene} provides a challenging experiment (the Suzanne scene) featuring a long sequence of smoke interacting with complex obstacles. There, we comprehensively evaluate our method against baselines using quantitative velocity metrics as well as re-simulation tasks in both training and novel testing views. \rv{Finally, Sec.~\ref{sec:ablation} presents ablation studies. Furthermore, additional evaluations regarding obstacle interactions and robustness in highly turbulent scenarios, along with implementation details, are deferred to the supplementary.}
}

\subsection{Comparisons on Standard Benchmarks}
\label{sec:comp_standard}

We evaluate our method against all baselines on the ScalarSyn and ScalarReal datasets. 

We first evaluate transport consistency through re-simulation. As shown in Fig.~\ref{fig:resimSyn} and Fig.~\ref{fig:resimReal}, baseline methods often suffer from noticeable density drift and dissipation over long sequences, failing to preserve the smoke's shape. Our method, conversely, maintains sharp structural details and correct advection trajectories even at late timestamps. As shown in Table~\ref{tbl:std_bench}, this qualitative observation is also mirrored by our quantitative metrics: we consistently outperform all baselines in PSNR, SSIM, and LPIPS. We attribute this robustness to our Sliding Window strategy, which propagates gradients from future observations back to earlier frames. This mechanism leverages temporal foresight to lock the velocity field to the smoke's true Lagrangian trajectory, ensuring long-term consistency.

\begin{figure*}[htbp] 
    \centering
    \setlength{\imagewidth}{0.32\textwidth} 
    
    \begin{minipage}[b]{\imagewidth}
        \centering
        \includegraphics[width=\linewidth]{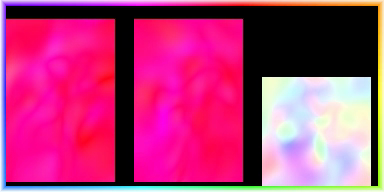}
        \vspace{-1.9em} 
        \caption*{PINF}
    \end{minipage}%
    \begin{minipage}[b]{\imagewidth}
        \centering
        \includegraphics[width=\linewidth]{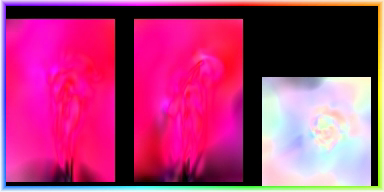}
        \vspace{-1.9em}
        \caption*{PICT}
    \end{minipage}%
    \begin{minipage}[b]{\imagewidth}
        \centering
        \includegraphics[width=\linewidth]{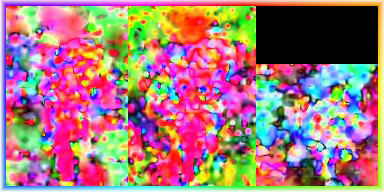}
        \vspace{-1.9em}
        \caption*{HyFluid}
    \end{minipage}
    
    
    \begin{minipage}[b]{\linewidth}
        \centering
        \begin{minipage}[b]{\imagewidth}
            \centering
            \includegraphics[width=\linewidth]{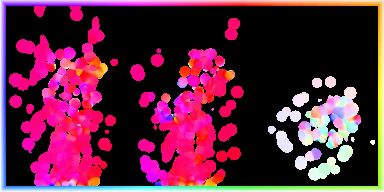}
            \vspace{-1.9em}
            \caption*{FluidNexus}
        \end{minipage}%
        \begin{minipage}[b]{\imagewidth}
            \centering
            \includegraphics[width=\linewidth]{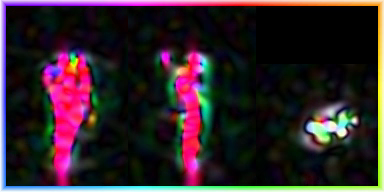}
            \vspace{-1.9em}
            \caption*{Ours}
        \end{minipage}
    \end{minipage}
    \caption{Visualization of reconstructed velocity fields on the ScalarReal dataset. Compared to baselines which often produce noisy or chaotic motion artifacts in background regions, our method generates a smooth and physically plausible velocity field. The clean flow structure and clear boundaries demonstrate the robustness of our DFK representation on real-world data.}
    \label{fig:velcmpReal}
\end{figure*}

Next, we examine the physical validity of the reconstructed velocity fields. As illustrated in Fig.~\ref{fig:velcmpSyn} and Fig.~\ref{fig:velcmpReal}, our method produces clean flow structures that closely align with the Ground Truth. In comparison, PINF and PICT tend to produce overly smoothed fields lacking fine details, whereas HyFluid exhibits significant chaotic noise. FluidNexus, limited by its discrete PBF representation, often generates discontinuous, spotty velocity fields. Quantitatively, As shown in Table~\ref{tbl:std_bench}, our method achieves the lowest Velocity MSE and highest Cosine Similarity on ScalarSyn, indicating superior accuracy in recovered dynamics. Furthermore, thanks to the DFK representation, we maintain a near-zero divergence across all datasets, which is often orders of magnitude lower than baselines. This confirms that integrating a strictly divergence-free basis by construction is far more effective than optimizing soft physical penalties.

\begin{figure}[bp]
    \centering
    
    \begin{minipage}[t]{0.48\linewidth}
        \centering
        \includegraphics[width=\linewidth]{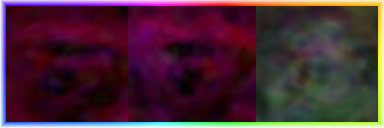} \\
        \vspace{-0.4em} 
        {\small PINF}
    \end{minipage}
    \begin{minipage}[t]{0.48\linewidth}
        \centering
        \includegraphics[width=\linewidth]{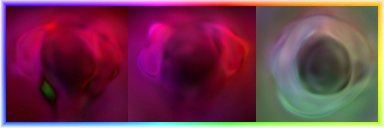} \\
        \vspace{-0.4em}
        {\small PICT}
    \end{minipage}

    \vspace{0.2em} 

    \begin{minipage}[t]{0.48\linewidth}
        \centering
        \includegraphics[width=\linewidth]{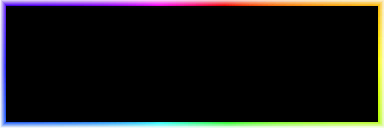} \\
        \vspace{-0.4em}
        {\small FluidNexus}
    \end{minipage}
    \begin{minipage}[t]{0.48\linewidth}
        \centering
        \includegraphics[width=\linewidth]{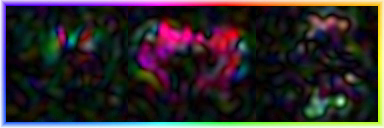} \\
        \vspace{-0.4em}
        {\small Ours}
    \end{minipage}

    \vspace{0.2em} 

    \begin{minipage}[t]{0.48\linewidth}
        \centering
        \includegraphics[width=\linewidth]{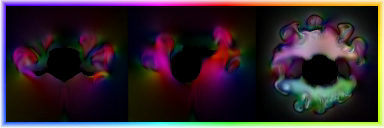} \\
        \vspace{-0.4em}
        {\small GT}
    \end{minipage}

    \vspace{-6pt}
    \caption{\rv{Visualization of reconstructed velocity fields on the Suzanne scene. While PINF and PICT can reconstruct plausible overall flow patterns, our method recovers flow structures that more closely align with the Ground Truth. Notably, the velocity field produced by FluidNexus collapses to near-zero, as its extensively hard-coded settings severely limit its generalization to this novel setup.}}
    \label{fig:supp_vel_suzanne}
\end{figure}
\begin{figure}[bp]
    \centering
    
    \setlength{\imagewidth}{0.16\textwidth}
    
    \def\synHeight{140pt} 
    
    \newcommand{\formattedgraphics}[5]{%
        \begin{tikzpicture}[spy using outlines={rectangle, magnification=2, connect spies}]
          \fill[black] (0, 5pt) rectangle (\imagewidth, \synHeight);
          
          \clip (0, 5pt) rectangle (\imagewidth, \synHeight);
          
          \node[anchor=south west, inner sep=0] at (0,0){\includegraphics[width=\imagewidth]{#1}};
          
          \node[anchor=north west,text=white] at (.02\imagewidth, \synHeight - 5pt){\sffamily\footnotesize\textbf{#2}};
          
          \node[anchor=north west,text=white] at (.02\imagewidth, \synHeight - 15pt){\sffamily\scriptsize #3};
          \node[anchor=north west,text=white] at (.02\imagewidth, \synHeight - 22pt){\sffamily\scriptsize #4};
          \node[anchor=north west,text=white] at (.02\imagewidth, \synHeight - 29pt){\sffamily\scriptsize #5};
        \end{tikzpicture}%
    }
    \formattedgraphics{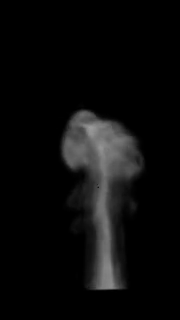}{Ours ($\gamma = 0.1$)}{PSNR (Avg) 42.72}{SSIM (Avg) 0.9862}{LPIPS (Avg) 0.03982}%
    \hspace{-0.18cm}%
    \formattedgraphics{fig/scalarsyn_resim/DFKGS_final.png}{Ours ($\gamma = 0.9$)}{PSNR (Avg) \textbf{46.50}}{SSIM (Avg) \textbf{0.9930}}{LPIPS (Avg) \textbf{0.02581}}%
    \hspace{-0.18cm}%
    \formattedgraphics{fig/scalarsyn_resim/gt_030.png}{Ground Truth}{}{}{}%
    \\
    \vspace{-6pt}
    \caption{\rv{Ablation study on the Sliding Window strategy. We compare the re-simulation quality of our method under different temporal discount factors: $\gamma = 0.9$ (our default long-range setting) versus $\gamma = 0.1$ (approximating short-sighted optimization). While the short-sighted model captures the general motion, it suffers from degradation in fine structural details and transport accuracy, as reflected by the lower PSNR, SSIM, and higher LPIPS scores. In contrast, our full model with $\gamma = 0.9$ effectively utilizes long-range temporal foresight to maintain high-fidelity consistency with the Ground Truth.}}
    \label{fig:ablation_sliding}
\end{figure}

\begin{figure*}[htbp]
    \centering
    
    \setlength{\imagewidth}{0.20\textwidth} 
    \def\trainHeight{110pt} 
    
    \newcommand{\formattedgraphicsTrain}[3]{%
        \begin{tikzpicture}[spy using outlines={rectangle, magnification=2, connect spies}]
          \fill[black] (0, 10pt) rectangle (\imagewidth, \trainHeight);
          \clip (0, 10pt) rectangle (\imagewidth, \trainHeight);
          \node[anchor=south west, inner sep=0] at (0,0){\includegraphics[width=\imagewidth]{#1}};
          \node[anchor=north west,text=white] at (.02\imagewidth, \trainHeight - 5pt){\sffamily\footnotesize\textbf{#2}};
          \node[anchor=north west,text=white] at (.02\imagewidth, \trainHeight - 15pt){\sffamily\scriptsize #3};
        \end{tikzpicture}%
    }

    \formattedgraphicsTrain{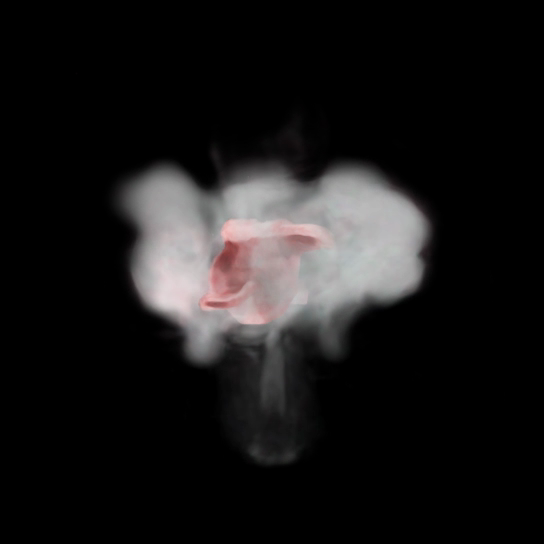}{PINF}{}%
    \hspace{-0.18cm}%
    \formattedgraphicsTrain{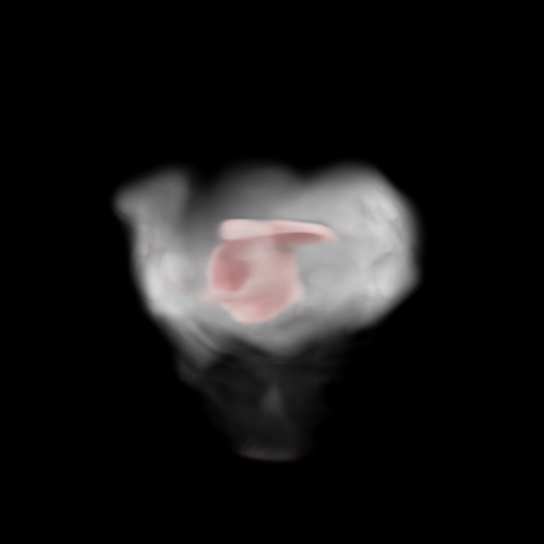}{PICT}{}%
    \hspace{-0.18cm}%
    \formattedgraphicsTrain{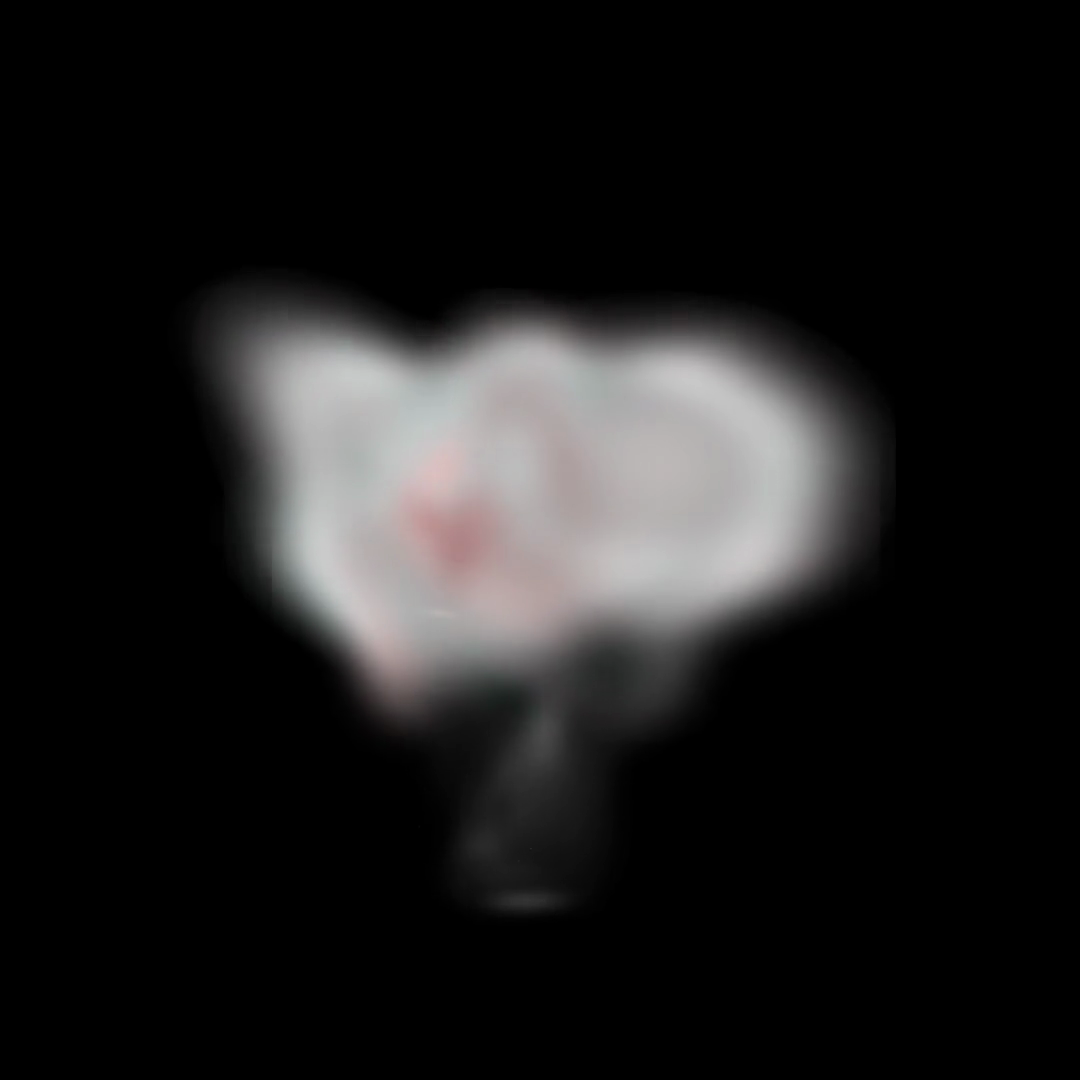}{FluidNexus}{}%
    \hspace{-0.18cm}%
    \formattedgraphicsTrain{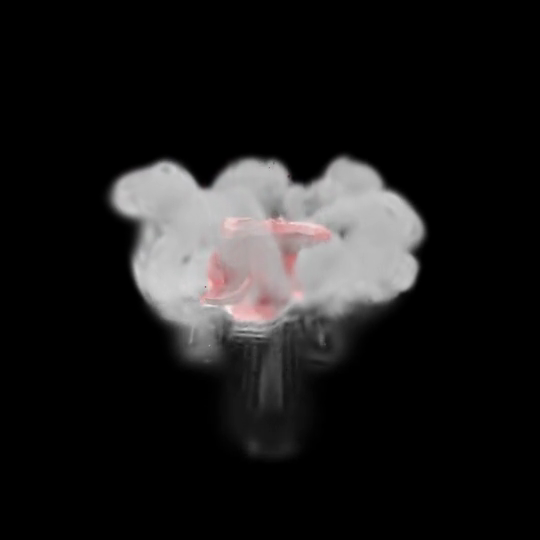}{Ours}{}%
    \hspace{-0.18cm}%
    \formattedgraphicsTrain{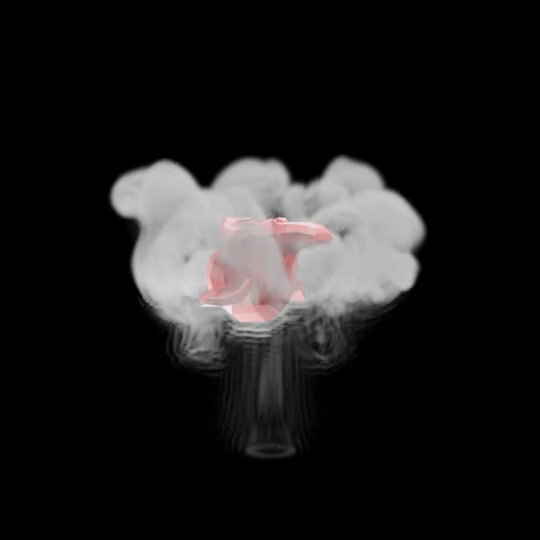}{Ground Truth}{}%
    \\
    \caption{\rv{Comparison of training-view re-simulation results on the Suzanne scene. While baseline methods struggle with severe density drift and structural degradation over the long duration, our method robustly handles the obstacle interaction, maintaining sharp structural details and accurate advection trajectories that closely match the Ground Truth.}}
    \label{fig:supp_resim_suzanne_train}
\end{figure*}

\begin{figure*}[htbp]
    \vspace{0.5em}
    \centering
    
    \setlength{\imagewidth}{0.20\textwidth} 
    \def\testHeight{110pt} 
    
    \newcommand{\formattedgraphicsTest}[3]{%
        \begin{tikzpicture}[spy using outlines={rectangle, magnification=2, connect spies}]
          \fill[black] (0, 10pt) rectangle (\imagewidth, \testHeight);
          \clip (0, 10pt) rectangle (\imagewidth, \testHeight);
          \node[anchor=south west, inner sep=0] at (0,0){\includegraphics[width=\imagewidth]{#1}};
          \node[anchor=north west,text=white] at (.02\imagewidth, \testHeight - 5pt){\sffamily\footnotesize\textbf{#2}};
          \node[anchor=north west,text=white] at (.02\imagewidth, \testHeight - 15pt){\sffamily\scriptsize #3};
        \end{tikzpicture}%
    }

    \formattedgraphicsTest{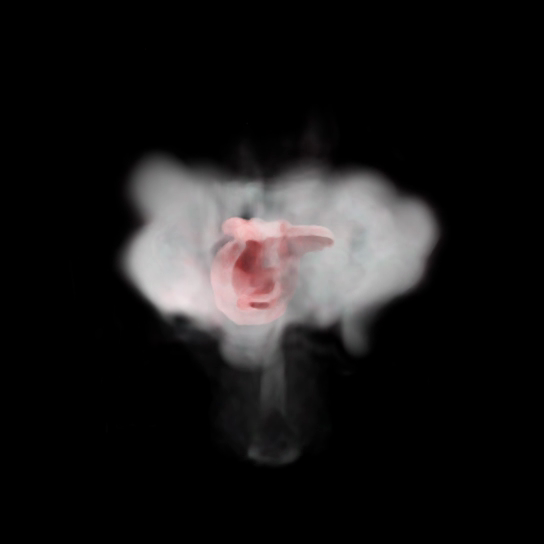}{PINF}{}%
    \hspace{-0.18cm}%
    \formattedgraphicsTest{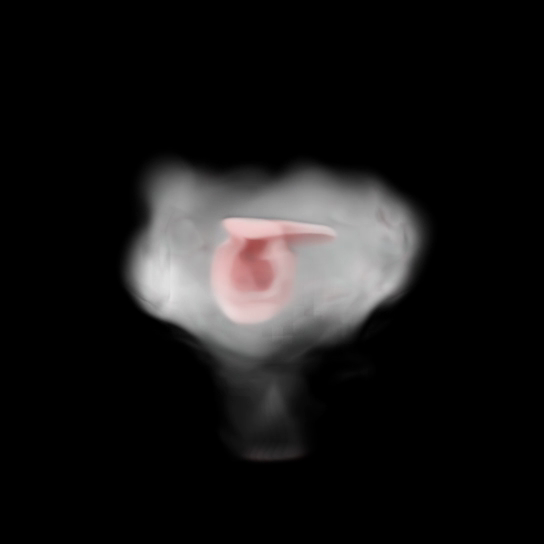}{PICT}{}%
    \hspace{-0.18cm}%
    \formattedgraphicsTest{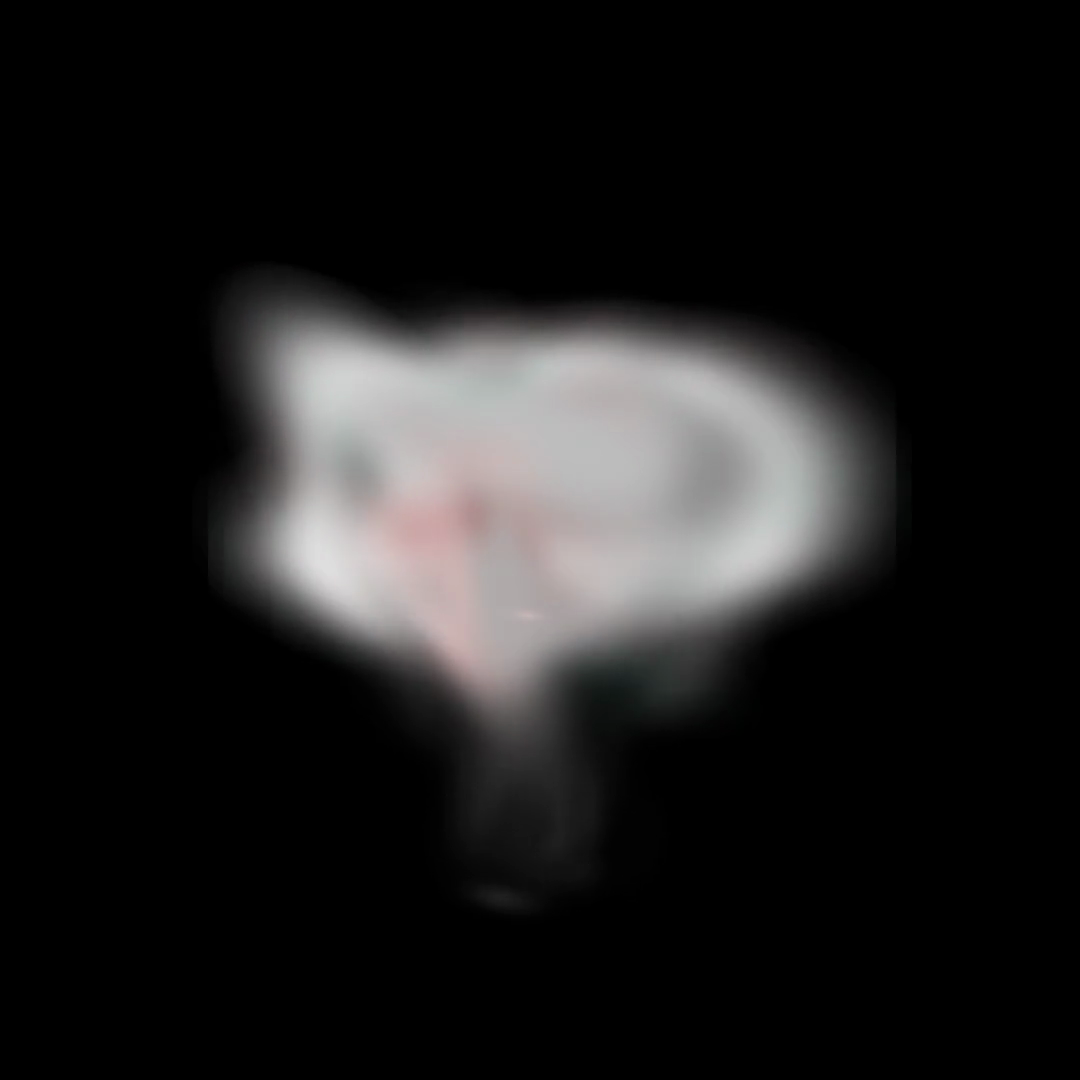}{FluidNexus}{}%
    \hspace{-0.18cm}%
    \formattedgraphicsTest{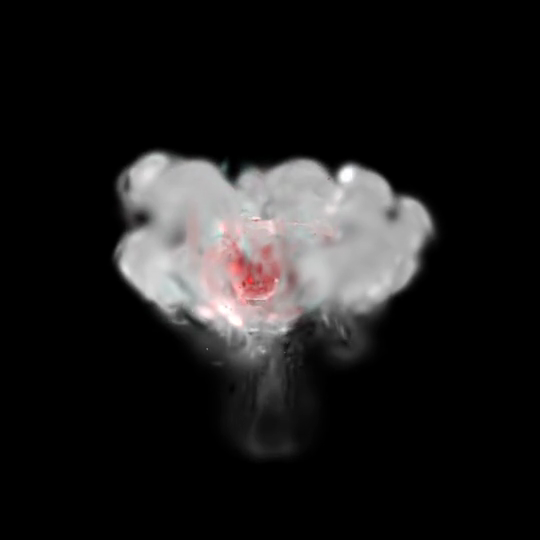}{Ours}{}%
    \hspace{-0.18cm}%
    \formattedgraphicsTest{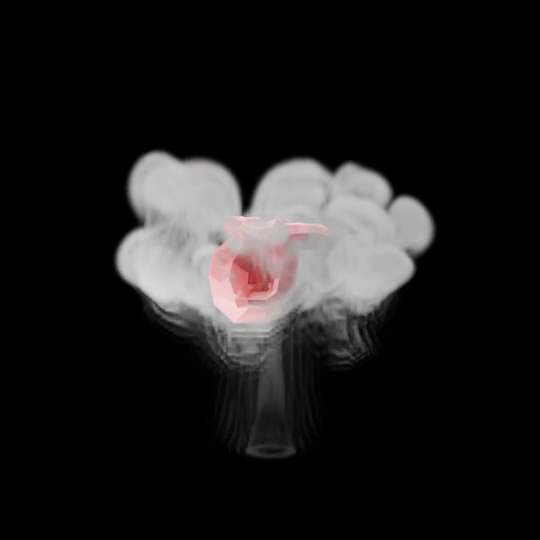}{Ground Truth}{}%
    \\
    \caption{\rv{Comparison of novel test-view re-simulation results on the Suzanne scene. By rendering the advected smoke from completely unseen camera angles, we evaluate the 3D physical and structural consistency of the reconstructed fields. While baseline methods exhibit clear volumetric artifacts and incorrect density distributions, our framework synthesizes structurally plausible novel views with superior PSNR and SSIM. We also observe that our method yields slightly higher LPIPS scores. We attribute this finding to a degree of overfitting to the training views, which can occasionally compromise fine textural details when rendered from unseen perspectives.}}
    \label{fig:supp_resim_suzanne_test}
\end{figure*}

\rv{
\subsection{Evaluation on Suzanne Scene}
\label{sec:supp_complex_scene}

To comprehensively evaluate our method's robustness over extended periods and its capability to handle intricate boundary conditions, we introduce the Suzanne scene. This scene simulates a rising plume of smoke interacting with the complex geometry of a monkey head model (Suzanne in Blender~\cite{blender}), spanning a long sequence of 10 seconds.

We benchmark our framework against PINF, PICT, and FluidNexus. Note that HyFluid is excluded from this evaluation because it does not support the reconstruction of colored scene. Additionally, we found that FluidNexus struggles to generalize to this new scenario. Due to its extensively hard-coded settings, such as fixed priors on initial smoke colors, FluidNexus fails entirely to drive the flow optimization, producing a degenerate, near-zero velocity field (as clearly evident in Fig.~\ref{fig:supp_vel_suzanne}).

Consistent with previous results, our method demonstrates superior performance in both physical validity and transport consistency on this scene, as summarized in Table~\ref{tbl:supp_suzanne_bench}. Quantitatively, our reconstructed velocity fields achieve the best results across divergence (Div), mean squared error (MSE-$v$), and cosine similarity (Cos-$v$). As visualized in Figure~\ref{fig:supp_vel_suzanne}, our method reconstructs a velocity field that most closely resembles the Ground Truth, even in the presence of complex obstacles. For the training-view re-simulation task, our approach consistently outperforms all evaluated baselines in PSNR, SSIM, and LPIPS. As illustrated in Figure~\ref{fig:supp_resim_suzanne_train}, our method robustly preserves sharp structural details over the long duration, whereas other approaches suffer from severe density drift.

We also evaluate the re-simulation performance on novel test views to assess the generalization of the reconstructed physics and representations, as quantitatively reported in Table~\ref{tbl:supp_suzanne_bench} and visually compared in Figure~\ref{fig:supp_resim_suzanne_test}. In the novel test view, our method continues to surpass existing approaches in terms of PSNR and SSIM. We note that our method yields slightly higher LPIPS scores on the test views. 
We attribute this to the Gaussian representation's inherent tendency to overfit training views. This can compromise high-frequency textural details when rendered from unseen perspectives, presenting an interesting direction for future improvement.
}


\begin{table*}[tbp]
\caption{\rv{Quantitative comparison on the long-duration Suzanne scene. We evaluate across three aspects: physical validity (Velocity Metrics), training-view visual fidelity (Train View Re-Simulation), and novel-view generalization (Test View Re-Simulation). Note that HyFluid is excluded as it does not support colored scene reconstruction. FluidNexus struggles to reconstruct a physically meaningful velocity field and collapses to near zero velocity in this scenario due to its extensively hard-coded settings (e.g., fixed assumptions about initial smoke color). Our method achieves the best performance across all velocity and training-view metrics, and maintains superior PSNR and SSIM on novel test views, demonstrating robust transport consistency and generalization.}}
\vspace{-0.5em}
\centering
\small
\begin{tabular}{lccccccccc}
\toprule
\multirow{2}{*}{Model} & \multicolumn{3}{c}{Velocity Metrics} & \multicolumn{3}{c}{Train View Re-Simulation} & \multicolumn{3}{c}{Test View Re-Simulation} \\
\cmidrule(lr){2-4} \cmidrule(lr){5-7} \cmidrule(lr){8-10}
 & Div$\downarrow$ & MSE-$v$$\downarrow$ & Cos-$v$$\uparrow$ & PSNR$\uparrow$ & SSIM$\uparrow$ & LPIPS$\downarrow$ & PSNR$\uparrow$ & SSIM$\uparrow$ & LPIPS$\downarrow$ \\
\midrule
PINF       & 0.000260 & 0.02136 & 0.1947 & 27.80 & 0.9241 & 0.1199 & 26.91 & 0.9221 & 0.1236 \\
PICT       & 0.000665 & 0.05257 & 0.1679 & 24.34 & 0.9213 & 0.1115 & 23.63 & 0.9168 & \textbf{0.1171} \\
FluidNexus & \multicolumn{3}{c}{Collapses to zero} & 25.18 & 0.9231 & 0.1292 & 24.21 & 0.9178 & 0.1331 \\
Ours       & \textbf{0.000014} & \textbf{0.02120} & \textbf{0.2087} & \textbf{39.54} & \textbf{0.9731} & \textbf{0.1066} & \textbf{29.36} & \textbf{0.9383} & 0.1392 \\
\bottomrule
\end{tabular}
\label{tbl:supp_suzanne_bench}
\end{table*}

\rv{
\subsection{Ablation Study}
\label{sec:ablation}

We conduct ablation studies on the ScalarSyn dataset to validate the effect of our Sliding Window strategy for transport consistency and the impact of physical regularization terms on velocity accuracy.

\paragraph{Effect of Sliding Window Strategy.}
To assess the importance of long-range gradient propagation, we compare our default setting where the temporal discount factor $\gamma$ is 0.9 against a variant with a reduced factor of $\gamma = 0.1$. This lower value severely limits the temporal horizon, approximating a short-sighted optimization. As shown in Fig.~\ref{fig:ablation_sliding}, reducing the temporal foresight leads to a consistent degradation in re-simulation quality across all metrics, including PSNR, SSIM, and LPIPS. This confirms that propagating gradients from future observations is essential for maintaining high-fidelity transport consistency and recovering accurate flow trajectories.

\begin{figure}[bp]
    \centering
    \begin{minipage}[t]{\linewidth}
        \centering
        \begin{minipage}[t]{0.5\linewidth}
            \includegraphics[width=\linewidth]{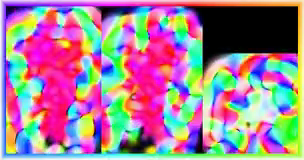}
            \vspace{-20pt}
            \caption*{w/o $\mathcal{L}_{\text{vor}} $ and $\mathcal{L}_{\text{reg}}$ (MSE-$v$ = $143.7$)}
            \includegraphics[width=\linewidth]{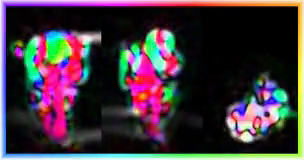}
            \vspace{-20pt}
            \caption*{w/o $\mathcal{L}_{\text{vor}}$ (MSE-$v$ = $0.1548$)}

        \end{minipage}%
        \hfill
        \begin{minipage}[t]{0.5\linewidth}
            \centering
            \includegraphics[width=\textwidth]{fig/scalarsyn/DFKGS_028.png}
            \vspace{-20pt}
            \caption*{Full (MSE-$v$ = $\mathbf{0.1116}$)}
            
            \includegraphics[width=\textwidth]{fig/scalarsyn/gt_028.png}
            \vspace{-20pt}
            \caption*{{Ground Truth}}
        \end{minipage}
    \end{minipage}
    \vspace{-6pt}
    \caption{\rv{Ablation study on physical regularization terms. We compare velocity field reconstructions under different loss configurations: (Top-Left) without any physical regularization, (Bottom-Left) with only kinetic energy regularization $\mathcal{L}_{\text{reg}}$, and (Top-Right) our full model including vorticity transport loss $\mathcal{L}_{\text{vor}}$. As shown, removing all regularizers leads to significant background noise and high Velocity MSE. Introducing $\mathcal{L}_{\text{reg}}$ suppresses spurious motion in empty regions, drastically reducing the error. Finally, incorporating $\mathcal{L}_{\text{vor}}$ further refines the flow structure to match the Ground Truth (Bottom-Right), demonstrating that these physics-informed losses are crucial for recovering accurate flow dynamics beyond basic divergence-free representation.}}
    \label{fig:ablation_physics}
\end{figure}

\paragraph{Effect of Physical Regularizations.}
While the DFK representation guarantees divergence-free motion by construction, we incorporate additional loss terms to further refine the solution. We evaluate the impact of the kinetic energy regularization $\mathcal{L}_{\text{reg}}$ and the vorticity transport loss $\mathcal{L}_{\text{vor}}$ by training variants without them. As illustrated in Fig.~\ref{fig:ablation_physics}, removing both terms results in a higher velocity Mean Squared Error. Adding the kinetic energy regularization $\mathcal{L}_{\text{reg}}$ effectively suppresses noise in empty regions. Incorporating the vorticity transport loss $\mathcal{L}_{\text{vor}}$ further improves the flow accuracy, yielding the best performance. This demonstrates that while DFK ensures kinematic validity through mass conservation, these physics-informed losses are valuable for recovering the correct dynamics governed by momentum conservation.
}

\section{Conclusion}

In this paper, we presented a novel framework for reconstructing high-fidelity fluid velocity fields from sparse multi-view videos. By integrating the efficiency of 3D Gaussian Splatting with the physical rigor of Divergence-Free Kernels, we enforce strict incompressibility by construction. Furthermore, our Sliding Window optimization strategy successfully enforces long-term Lagrangian consistency while maintaining computational tractability. This mechanism enables the propagation of gradients over meaningful temporal horizons, recovering accurate flow dynamics that are both visually consistent with the input observations and physically plausible.

Despite these advancements, our method is subject to several limitations that point toward exciting avenues for future research.

First, our current formulation is primarily designed for divergence-free, incompressible fluids (e.g., smoke).  \rv{
Because it relies on DFKs, the method does not readily extend to compressible flows, and it can be challenging to accurately capture highly nonlinear and chaotic fluid dynamics, such as shocks. Other phenomena like free-surface liquids, elastic bodies, or cloth are also not covered in the present work.
} Extending our Lagrangian-based sliding window framework to recover the dynamics and physical parameters of these diverse materials \rv{and complex flow regimes} remains a promising direction.
Second, while DFK guarantees mass conservation (divergence-free), the momentum conservation is only encouraged via soft physics-informed losses. Consequently, discrepancies between the reconstructed velocity and the ground truth may persist in complex scenarios. Developing a velocity parameterization that naturally satisfies momentum transport by construction could lead to improvements in reconstruction accuracy.
Third, although our method achieves a substantial speedup compared to implicit neural representations, it is not yet capable of real-time reconstruction. \tnx{Specifically, our total training time typically ranges from 30 minutes to 3 hours depending on scene complexity, whereas implicit representations often require over ten hours. Detailed per-scene timings are provided in the supplementary.} Bridging the gap between offline optimization and real-time inference is crucial for deploying such techniques in time-sensitive applications like interactive aerodynamic analysis. We believe that further optimizing the computational graph or exploring feed-forward prediction models could be key to unlocking this potential.


\begin{acks}
This work was supported by the National Natural Science Foundation of China (Project No. 62595772).
\end{acks}



{\small
	\bibliographystyle{ACM-Reference-Format}
	\bibliography{egbib}


\begin{thebibliography}{52}


\ifx \showCODEN    \undefined \def \showCODEN     #1{\unskip}     \fi
\ifx \showDOI      \undefined \def \showDOI       #1{#1}\fi
\ifx \showISBNx    \undefined \def \showISBNx     #1{\unskip}     \fi
\ifx \showISBNxiii \undefined \def \showISBNxiii  #1{\unskip}     \fi
\ifx \showISSN     \undefined \def \showISSN      #1{\unskip}     \fi
\ifx \showLCCN     \undefined \def \showLCCN      #1{\unskip}     \fi
\ifx \shownote     \undefined \def \shownote      #1{#1}          \fi
\ifx \showarticletitle \undefined \def \showarticletitle #1{#1}   \fi
\ifx \showURL      \undefined \def \showURL       {\relax}        \fi
\providecommand\bibfield[2]{#2}
\providecommand\bibinfo[2]{#2}
\providecommand\natexlab[1]{#1}
\providecommand\showeprint[2][]{arXiv:#2}

\bibitem[Atcheson et~al\mbox{.}(2009)]%
        {Atcheson2008OEF}
\bibfield{author}{\bibinfo{person}{Bradley Atcheson}, \bibinfo{person}{Wolfgang Heidrich}, {and} \bibinfo{person}{Ivo Ihrke}.} \bibinfo{year}{2009}\natexlab{}.
\newblock \showarticletitle{An evaluation of optical flow algorithms for background oriented schlieren imaging}.
\newblock \bibinfo{journal}{\emph{Experiments in Fluids}} \bibinfo{volume}{46}, \bibinfo{number}{3} (\bibinfo{date}{01 Mar} \bibinfo{year}{2009}), \bibinfo{pages}{467--476}.
\newblock
\showISSN{1432-1114}
\urldef\tempurl%
\url{https://doi.org/10.1007/s00348-008-0572-7}
\showDOI{\tempurl}


\bibitem[Barron et~al\mbox{.}(2021)]%
        {barron-ICCV2021-mipnerf}
\bibfield{author}{\bibinfo{person}{Jonathan~T Barron}, \bibinfo{person}{Ben Mildenhall}, \bibinfo{person}{Matthew Tancik}, \bibinfo{person}{Peter Hedman}, \bibinfo{person}{Ricardo Martin-Brualla}, {and} \bibinfo{person}{Pratul~P Srinivasan}.} \bibinfo{year}{2021}\natexlab{}.
\newblock \showarticletitle{Mip-nerf: A multiscale representation for anti-aliasing neural radiance fields}. In \bibinfo{booktitle}{\emph{Proceedings of the IEEE/CVF International Conference on Computer Vision}}. \bibinfo{pages}{5855--5864}.
\newblock


\bibitem[Bi et~al\mbox{.}(2023)]%
        {bi2023accurate}
\bibfield{author}{\bibinfo{person}{K Bi}, \bibinfo{person}{L Xie}, \bibinfo{person}{H Zhang}, \bibinfo{person}{X Chen}, \bibinfo{person}{X Gu}, {and} \bibinfo{person}{Q Tian}.} \bibinfo{year}{2023}\natexlab{}.
\newblock \showarticletitle{Accurate medium-range global weather forecasting with 3D neural networks.}
\newblock \bibinfo{journal}{\emph{Nature}} \bibinfo{volume}{619}, \bibinfo{number}{7970} (\bibinfo{year}{2023}), \bibinfo{pages}{533--538}.
\newblock


\bibitem[{Blender}(2011)]%
        {blender}
\bibfield{author}{\bibinfo{person}{{Blender}}.} \bibinfo{year}{2011}\natexlab{}.
\newblock \bibinfo{title}{(CC) Blender Foundation}.
\newblock \bibinfo{howpublished}{\url{https://www.blender.org/}}.
\newblock
\newblock
\shownote{Online}.


\bibitem[Bridson(2015)]%
        {bridson2015fluid}
\bibfield{author}{\bibinfo{person}{Robert Bridson}.} \bibinfo{year}{2015}\natexlab{}.
\newblock \bibinfo{booktitle}{\emph{Fluid simulation for computer graphics}}.
\newblock \bibinfo{publisher}{CRC press}.
\newblock


\bibitem[Cao and Zhang(2025)]%
        {cao2025analysis}
\bibfield{author}{\bibinfo{person}{Wenbo Cao} {and} \bibinfo{person}{Weiwei Zhang}.} \bibinfo{year}{2025}\natexlab{}.
\newblock \showarticletitle{An analysis and solution of ill-conditioning in physics-informed neural networks}.
\newblock \bibinfo{journal}{\emph{J. Comput. Phys.}}  \bibinfo{volume}{520} (\bibinfo{year}{2025}), \bibinfo{pages}{113494}.
\newblock


\bibitem[Chang et~al\mbox{.}(2021)]%
        {chang2021curl}
\bibfield{author}{\bibinfo{person}{Jumyung Chang}, \bibinfo{person}{Ruben Partono}, \bibinfo{person}{Vinicius~C Azevedo}, {and} \bibinfo{person}{Christopher Batty}.} \bibinfo{year}{2021}\natexlab{}.
\newblock \showarticletitle{Curl-flow: Boundary-respecting pointwise incompressible velocity interpolation for grid-based fluids}.
\newblock \bibinfo{journal}{\emph{arXiv preprint arXiv:2104.00867}} (\bibinfo{year}{2021}).
\newblock


\bibitem[Chen et~al\mbox{.}(2023)]%
        {chen2023implicit}
\bibfield{author}{\bibinfo{person}{Honglin Chen}, \bibinfo{person}{Rundi Wu}, \bibinfo{person}{Eitan Grinspun}, \bibinfo{person}{Changxi Zheng}, {and} \bibinfo{person}{Peter~Yichen Chen}.} \bibinfo{year}{2023}\natexlab{}.
\newblock \showarticletitle{Implicit neural spatial representations for time-dependent pdes}. In \bibinfo{booktitle}{\emph{International Conference on Machine Learning}}. PMLR, \bibinfo{pages}{5162--5177}.
\newblock


\bibitem[Chu et~al\mbox{.}(2022)]%
        {chu2022pinf}
\bibfield{author}{\bibinfo{person}{Mengyu Chu}, \bibinfo{person}{Lingjie Liu}, \bibinfo{person}{Quan Zheng}, \bibinfo{person}{Erik Franz}, \bibinfo{person}{Hans-Peter Seidel}, \bibinfo{person}{Christian Theobalt}, {and} \bibinfo{person}{Rhaleb Zayer}.} \bibinfo{year}{2022}\natexlab{}.
\newblock \showarticletitle{Physics informed neural fields for smoke reconstruction with sparse data}.
\newblock \bibinfo{journal}{\emph{ACM Transactions on Graphics (TOG)}} \bibinfo{volume}{41}, \bibinfo{number}{4} (\bibinfo{year}{2022}), \bibinfo{pages}{1--14}.
\newblock


\bibitem[Cottet(2001)]%
        {cottet2001vortex}
\bibfield{author}{\bibinfo{person}{Georges-Henri Cottet}.} \bibinfo{year}{2001}\natexlab{}.
\newblock \showarticletitle{Vortex methods: theory and practice}.
\newblock \bibinfo{journal}{\emph{Measurement Science and Technology}} (\bibinfo{year}{2001}).
\newblock


\bibitem[Du et~al\mbox{.}(2025)]%
        {du2025gaussfluids}
\bibfield{author}{\bibinfo{person}{F. Du}, \bibinfo{person}{Y. Zhang}, \bibinfo{person}{Y. Ji}, \bibinfo{person}{X. Wang}, \bibinfo{person}{C. Yao}, \bibinfo{person}{J. Kosinka}, \bibinfo{person}{S. Frey}, \bibinfo{person}{A. Telea}, {and} \bibinfo{person}{X. Ban}.} \bibinfo{year}{2025}\natexlab{}.
\newblock \showarticletitle{GaussFluids: Reconstructing Lagrangian Fluid Particles from Videos via Gaussian Splatting}. In \bibinfo{booktitle}{\emph{Pacific Graphics}}.
\newblock


\bibitem[Duan et~al\mbox{.}(2024)]%
        {duan20244d}
\bibfield{author}{\bibinfo{person}{Yuanxing Duan}, \bibinfo{person}{Fangyin Wei}, \bibinfo{person}{Qiyu Dai}, \bibinfo{person}{Yuhang He}, \bibinfo{person}{Wenzheng Chen}, {and} \bibinfo{person}{Baoquan Chen}.} \bibinfo{year}{2024}\natexlab{}.
\newblock \showarticletitle{4d-rotor gaussian splatting: towards efficient novel view synthesis for dynamic scenes}. In \bibinfo{booktitle}{\emph{ACM SIGGRAPH 2024 Conference Papers}}. \bibinfo{pages}{1--11}.
\newblock


\bibitem[Eckert et~al\mbox{.}(2019)]%
        {eckert2019scalarflow}
\bibfield{author}{\bibinfo{person}{Marie-Lena Eckert}, \bibinfo{person}{Kiwon Um}, {and} \bibinfo{person}{Nils Thuerey}.} \bibinfo{year}{2019}\natexlab{}.
\newblock \showarticletitle{ScalarFlow: a large-scale volumetric data set of real-world scalar transport flows for computer animation and machine learning}.
\newblock \bibinfo{journal}{\emph{ACM Transactions on Graphics (TOG)}} \bibinfo{volume}{38}, \bibinfo{number}{6} (\bibinfo{year}{2019}), \bibinfo{pages}{1--16}.
\newblock


\bibitem[Elsinga et~al\mbox{.}(2006)]%
        {elsinga2006tomographic}
\bibfield{author}{\bibinfo{person}{Gerrit~E Elsinga}, \bibinfo{person}{Fulvio Scarano}, \bibinfo{person}{Bernhard Wieneke}, {and} \bibinfo{person}{Bas~W van Oudheusden}.} \bibinfo{year}{2006}\natexlab{}.
\newblock \showarticletitle{Tomographic particle image velocimetry}.
\newblock \bibinfo{journal}{\emph{Experiments in fluids}} \bibinfo{volume}{41}, \bibinfo{number}{6} (\bibinfo{year}{2006}), \bibinfo{pages}{933--947}.
\newblock


\bibitem[Franz et~al\mbox{.}(2021)]%
        {franz2021global}
\bibfield{author}{\bibinfo{person}{Erik Franz}, \bibinfo{person}{Barbara Solenthaler}, {and} \bibinfo{person}{Nils Thuerey}.} \bibinfo{year}{2021}\natexlab{}.
\newblock \showarticletitle{Global Transport for Fluid Reconstruction with Learned Self-Supervision}. In \bibinfo{booktitle}{\emph{Proceedings of the IEEE/CVF Conference on Computer Vision and Pattern Recognition}}. \bibinfo{pages}{1632--1642}.
\newblock


\bibitem[Gao et~al\mbox{.}(2025)]%
        {gao2025fluidnexus}
\bibfield{author}{\bibinfo{person}{Yue Gao}, \bibinfo{person}{Hong-Xing Yu}, \bibinfo{person}{Bo Zhu}, {and} \bibinfo{person}{Jiajun Wu}.} \bibinfo{year}{2025}\natexlab{}.
\newblock \showarticletitle{FluidNexus: 3D fluid reconstruction and prediction from a single video}. In \bibinfo{booktitle}{\emph{Proceedings of the Computer Vision and Pattern Recognition Conference}}. \bibinfo{pages}{26091--26101}.
\newblock


\bibitem[Grant(1997)]%
        {grant1997particle}
\bibfield{author}{\bibinfo{person}{Ian Grant}.} \bibinfo{year}{1997}\natexlab{}.
\newblock \showarticletitle{Particle image velocimetry: a review}.
\newblock \bibinfo{journal}{\emph{Proceedings of the Institution of Mechanical Engineers, Part C: Journal of Mechanical Engineering Science}} \bibinfo{volume}{211}, \bibinfo{number}{1} (\bibinfo{year}{1997}), \bibinfo{pages}{55--76}.
\newblock


\bibitem[Gregson et~al\mbox{.}(2014)]%
        {gregson2014capture}
\bibfield{author}{\bibinfo{person}{James Gregson}, \bibinfo{person}{Ivo Ihrke}, \bibinfo{person}{Nils Thuerey}, {and} \bibinfo{person}{Wolfgang Heidrich}.} \bibinfo{year}{2014}\natexlab{}.
\newblock \showarticletitle{From capture to simulation: connecting forward and inverse problems in fluids}.
\newblock \bibinfo{journal}{\emph{ACM Transactions on Graphics (TOG)}} \bibinfo{volume}{33}, \bibinfo{number}{4} (\bibinfo{year}{2014}), \bibinfo{pages}{1--11}.
\newblock


\bibitem[Gu et~al\mbox{.}(2012)]%
        {gu2012compressive}
\bibfield{author}{\bibinfo{person}{Jinwei Gu}, \bibinfo{person}{Shree~K Nayar}, \bibinfo{person}{Eitan Grinspun}, \bibinfo{person}{Peter~N Belhumeur}, {and} \bibinfo{person}{Ravi Ramamoorthi}.} \bibinfo{year}{2012}\natexlab{}.
\newblock \showarticletitle{Compressive structured light for recovering inhomogeneous participating media}.
\newblock \bibinfo{journal}{\emph{IEEE transactions on pattern analysis and machine intelligence}} \bibinfo{volume}{35}, \bibinfo{number}{3} (\bibinfo{year}{2012}), \bibinfo{pages}{1--1}.
\newblock


\bibitem[Hong et~al\mbox{.}(2025)]%
        {hong2025physics}
\bibfield{author}{\bibinfo{person}{Haoqin Hong}, \bibinfo{person}{Ding Fan}, \bibinfo{person}{Fubin Dou}, \bibinfo{person}{Zhi-Li Zhou}, \bibinfo{person}{Haoran Sun}, \bibinfo{person}{Congcong Zhu}, {and} \bibinfo{person}{Jingrun Chen}.} \bibinfo{year}{2025}\natexlab{}.
\newblock \showarticletitle{Physics-Informed Deformable Gaussian Splatting: Towards Unified Constitutive Laws for Time-Evolving Material Field}.
\newblock \bibinfo{journal}{\emph{arXiv preprint arXiv:2511.06299}} (\bibinfo{year}{2025}).
\newblock


\bibitem[Ji et~al\mbox{.}(2013)]%
        {ji2013reconstructing}
\bibfield{author}{\bibinfo{person}{Yu Ji}, \bibinfo{person}{Jinwei Ye}, {and} \bibinfo{person}{Jingyi Yu}.} \bibinfo{year}{2013}\natexlab{}.
\newblock \showarticletitle{Reconstructing gas flows using light-path approximation}. In \bibinfo{booktitle}{\emph{Proceedings of the IEEE Conference on Computer Vision and Pattern Recognition}}. \bibinfo{pages}{2507--2514}.
\newblock


\bibitem[Jiang et~al\mbox{.}(2015)]%
        {jiang2015affine}
\bibfield{author}{\bibinfo{person}{Chenfanfu Jiang}, \bibinfo{person}{Craig Schroeder}, \bibinfo{person}{Andrew Selle}, \bibinfo{person}{Joseph Teran}, {and} \bibinfo{person}{Alexey Stomakhin}.} \bibinfo{year}{2015}\natexlab{}.
\newblock \showarticletitle{The affine particle-in-cell method}.
\newblock \bibinfo{journal}{\emph{ACM Transactions on Graphics (TOG)}} \bibinfo{volume}{34}, \bibinfo{number}{4} (\bibinfo{year}{2015}), \bibinfo{pages}{1--10}.
\newblock


\bibitem[Kerbl et~al\mbox{.}(2023)]%
        {kerbl2023gaussian_splatting}
\bibfield{author}{\bibinfo{person}{Bernhard Kerbl}, \bibinfo{person}{Georgios Kopanas}, \bibinfo{person}{Thomas Leimk{\"u}hler}, {and} \bibinfo{person}{George Drettakis}.} \bibinfo{year}{2023}\natexlab{}.
\newblock \showarticletitle{3d gaussian splatting for real-time radiance field rendering}.
\newblock \bibinfo{journal}{\emph{ACM Transactions on Graphics (ToG)}} \bibinfo{volume}{42}, \bibinfo{number}{4} (\bibinfo{year}{2023}), \bibinfo{pages}{1--14}.
\newblock


\bibitem[Kim et~al\mbox{.}(2019)]%
        {kim2019deep}
\bibfield{author}{\bibinfo{person}{Byungsoo Kim}, \bibinfo{person}{Vinicius~C Azevedo}, \bibinfo{person}{Nils Thuerey}, \bibinfo{person}{Theodore Kim}, \bibinfo{person}{Markus Gross}, {and} \bibinfo{person}{Barbara Solenthaler}.} \bibinfo{year}{2019}\natexlab{}.
\newblock \showarticletitle{Deep fluids: A generative network for parameterized fluid simulations}. In \bibinfo{booktitle}{\emph{Computer graphics forum}}, Vol.~\bibinfo{volume}{38}. Wiley Online Library, \bibinfo{pages}{59--70}.
\newblock


\bibitem[Luiten et~al\mbox{.}(2024)]%
        {luiten2024dynamic}
\bibfield{author}{\bibinfo{person}{Jonathon Luiten}, \bibinfo{person}{Georgios Kopanas}, \bibinfo{person}{Bastian Leibe}, {and} \bibinfo{person}{Deva Ramanan}.} \bibinfo{year}{2024}\natexlab{}.
\newblock \showarticletitle{Dynamic 3d gaussians: Tracking by persistent dynamic view synthesis}. In \bibinfo{booktitle}{\emph{2024 International Conference on 3D Vision (3DV)}}. IEEE, \bibinfo{pages}{800--809}.
\newblock


\bibitem[Lyu et~al\mbox{.}(2024)]%
        {lyu2024wavelet}
\bibfield{author}{\bibinfo{person}{Luan Lyu}, \bibinfo{person}{Xiaohua Ren}, \bibinfo{person}{Wei Cao}, \bibinfo{person}{Jian Zhu}, \bibinfo{person}{Enhua Wu}, {and} \bibinfo{person}{Zhi-Xin Yang}.} \bibinfo{year}{2024}\natexlab{}.
\newblock \showarticletitle{Wavelet potentials: An efficient potential recovery technique for pointwise incompressible fluids}. In \bibinfo{booktitle}{\emph{Computer Graphics Forum}}, Vol.~\bibinfo{volume}{43}. Wiley Online Library, \bibinfo{pages}{e15023}.
\newblock


\bibitem[Macklin and M{\"u}ller(2013)]%
        {macklin2013position}
\bibfield{author}{\bibinfo{person}{Miles Macklin} {and} \bibinfo{person}{Matthias M{\"u}ller}.} \bibinfo{year}{2013}\natexlab{}.
\newblock \showarticletitle{Position based fluids}.
\newblock \bibinfo{journal}{\emph{ACM Transactions on Graphics (TOG)}} \bibinfo{volume}{32}, \bibinfo{number}{4} (\bibinfo{year}{2013}), \bibinfo{pages}{1--12}.
\newblock


\bibitem[Mescheder et~al\mbox{.}(2019)]%
        {mescheder2019occupancy}
\bibfield{author}{\bibinfo{person}{Lars Mescheder}, \bibinfo{person}{Michael Oechsle}, \bibinfo{person}{Michael Niemeyer}, \bibinfo{person}{Sebastian Nowozin}, {and} \bibinfo{person}{Andreas Geiger}.} \bibinfo{year}{2019}\natexlab{}.
\newblock \showarticletitle{Occupancy networks: Learning 3d reconstruction in function space}. In \bibinfo{booktitle}{\emph{Proceedings of the IEEE/CVF conference on computer vision and pattern recognition}}. \bibinfo{pages}{4460--4470}.
\newblock


\bibitem[Mildenhall et~al\mbox{.}(2020)]%
        {mildenhall2020nerf}
\bibfield{author}{\bibinfo{person}{Ben Mildenhall}, \bibinfo{person}{Pratul~P. Srinivasan}, \bibinfo{person}{Matthew Tancik}, \bibinfo{person}{Jonathan~T. Barron}, \bibinfo{person}{Ravi Ramamoorthi}, {and} \bibinfo{person}{Ren Ng}.} \bibinfo{year}{2020}\natexlab{}.
\newblock \showarticletitle{NeRF: Representing Scenes as Neural Radiance Fields for View Synthesis}. In \bibinfo{booktitle}{\emph{ECCV}}.
\newblock


\bibitem[M{\"u}ller et~al\mbox{.}(2022)]%
        {muller-SIG2022-instantngp}
\bibfield{author}{\bibinfo{person}{Thomas M{\"u}ller}, \bibinfo{person}{Alex Evans}, \bibinfo{person}{Christoph Schied}, {and} \bibinfo{person}{Alexander Keller}.} \bibinfo{year}{2022}\natexlab{}.
\newblock \showarticletitle{Instant neural graphics primitives with a multiresolution hash encoding}.
\newblock \bibinfo{journal}{\emph{ACM Transactions on Graphics (ToG)}} \bibinfo{volume}{41}, \bibinfo{number}{4} (\bibinfo{year}{2022}), \bibinfo{pages}{1--15}.
\newblock


\bibitem[Nabizadeh et~al\mbox{.}(2024)]%
        {nabizadeh2024fluid}
\bibfield{author}{\bibinfo{person}{Mohammad~Sina Nabizadeh}, \bibinfo{person}{Ritoban Roy-Chowdhury}, \bibinfo{person}{Hang Yin}, \bibinfo{person}{Ravi Ramamoorthi}, {and} \bibinfo{person}{Albert Chern}.} \bibinfo{year}{2024}\natexlab{}.
\newblock \showarticletitle{Fluid Implicit Particles on Coadjoint Orbits}.
\newblock \bibinfo{journal}{\emph{arXiv preprint arXiv:2406.01936}} (\bibinfo{year}{2024}).
\newblock


\bibitem[Ni et~al\mbox{.}(2025)]%
        {ni2025representing}
\bibfield{author}{\bibinfo{person}{Xingyu Ni}, \bibinfo{person}{Jingrui Xing}, \bibinfo{person}{Xingqiao Li}, \bibinfo{person}{Bin Wang}, {and} \bibinfo{person}{Baoquan Chen}.} \bibinfo{year}{2025}\natexlab{}.
\newblock \showarticletitle{Representing Flow Fields with Divergence-Free Kernels for Reconstruction}.
\newblock \bibinfo{journal}{\emph{Proceedings of the ACM on Computer Graphics and Interactive Techniques}} \bibinfo{volume}{8}, \bibinfo{number}{4} (\bibinfo{year}{2025}), \bibinfo{pages}{1--21}.
\newblock


\bibitem[Park et~al\mbox{.}(2019)]%
        {park2019deepsdf}
\bibfield{author}{\bibinfo{person}{Jeong~Joon Park}, \bibinfo{person}{Peter Florence}, \bibinfo{person}{Julian Straub}, \bibinfo{person}{Richard Newcombe}, {and} \bibinfo{person}{Steven Lovegrove}.} \bibinfo{year}{2019}\natexlab{}.
\newblock \showarticletitle{{DeepSDF}: Learning continuous signed distance functions for shape representation}. In \bibinfo{booktitle}{\emph{Proceedings of the IEEE/CVF Conference on Computer Vision and Pattern Recognition}}. \bibinfo{pages}{165--174}.
\newblock


\bibitem[Peng et~al\mbox{.}(2020)]%
        {peng2020convolutional}
\bibfield{author}{\bibinfo{person}{Songyou Peng}, \bibinfo{person}{Michael Niemeyer}, \bibinfo{person}{Lars Mescheder}, \bibinfo{person}{Marc Pollefeys}, {and} \bibinfo{person}{Andreas Geiger}.} \bibinfo{year}{2020}\natexlab{}.
\newblock \showarticletitle{Convolutional occupancy networks}. In \bibinfo{booktitle}{\emph{Computer Vision--ECCV 2020: 16th European Conference, Glasgow, UK, August 23--28, 2020, Proceedings, Part III 16}}. Springer, \bibinfo{pages}{523--540}.
\newblock


\bibitem[Qiu et~al\mbox{.}(2024)]%
        {qiu2024neusmoke}
\bibfield{author}{\bibinfo{person}{Jiaxiong Qiu}, \bibinfo{person}{Ruihong Cen}, \bibinfo{person}{Zhong Li}, \bibinfo{person}{Han Yan}, \bibinfo{person}{Ming-Ming Cheng}, {and} \bibinfo{person}{Bo Ren}.} \bibinfo{year}{2024}\natexlab{}.
\newblock \showarticletitle{NeuSmoke: Efficient Smoke Reconstruction and View Synthesis with Neural Transportation Fields}. In \bibinfo{booktitle}{\emph{SIGGRAPH Asia 2024 Conference Papers}}. \bibinfo{pages}{1--12}.
\newblock


\bibitem[Raissi et~al\mbox{.}(2019)]%
        {RAISSI2019}
\bibfield{author}{\bibinfo{person}{M. Raissi}, \bibinfo{person}{P. Perdikaris}, {and} \bibinfo{person}{G.E. Karniadakis}.} \bibinfo{year}{2019}\natexlab{}.
\newblock \showarticletitle{Physics-informed neural networks: A deep learning framework for solving forward and inverse problems involving nonlinear partial differential equations}.
\newblock \bibinfo{journal}{\emph{J. Comput. Phys.}}  \bibinfo{volume}{378} (\bibinfo{year}{2019}), \bibinfo{pages}{686--707}.
\newblock
\showISSN{0021-9991}
\urldef\tempurl%
\url{https://doi.org/10.1016/j.jcp.2018.10.045}
\showDOI{\tempurl}


\bibitem[Rathore et~al\mbox{.}(2024)]%
        {rathore2024challenges}
\bibfield{author}{\bibinfo{person}{Pratik Rathore}, \bibinfo{person}{Weimu Lei}, \bibinfo{person}{Zachary Frangella}, \bibinfo{person}{Lu Lu}, {and} \bibinfo{person}{Madeleine Udell}.} \bibinfo{year}{2024}\natexlab{}.
\newblock \showarticletitle{Challenges in training PINNs: a loss landscape perspective}. In \bibinfo{booktitle}{\emph{Proceedings of the 41st International Conference on Machine Learning}}. \bibinfo{pages}{42159--42191}.
\newblock


\bibitem[Richter-Powell et~al\mbox{.}(2022)]%
        {richter2022neural}
\bibfield{author}{\bibinfo{person}{Jack Richter-Powell}, \bibinfo{person}{Yaron Lipman}, {and} \bibinfo{person}{Ricky~TQ Chen}.} \bibinfo{year}{2022}\natexlab{}.
\newblock \showarticletitle{Neural conservation laws: A divergence-free perspective}.
\newblock \bibinfo{journal}{\emph{Advances in Neural Information Processing Systems}}  \bibinfo{volume}{35} (\bibinfo{year}{2022}), \bibinfo{pages}{38075--38088}.
\newblock


\bibitem[Rosset et~al\mbox{.}(2023)]%
        {rosset2023interactive}
\bibfield{author}{\bibinfo{person}{Nicolas Rosset}, \bibinfo{person}{Guillaume Cordonnier}, \bibinfo{person}{Regis Duvigneau}, {and} \bibinfo{person}{Adrien Bousseau}.} \bibinfo{year}{2023}\natexlab{}.
\newblock \showarticletitle{Interactive design of 2D car profiles with aerodynamic feedback}. In \bibinfo{booktitle}{\emph{Computer Graphics Forum}}, Vol.~\bibinfo{volume}{42}. Wiley Online Library, \bibinfo{pages}{427--437}.
\newblock


\bibitem[Roy-Chowdhury et~al\mbox{.}(2024)]%
        {roy2024higher}
\bibfield{author}{\bibinfo{person}{Ritoban Roy-Chowdhury}, \bibinfo{person}{Tamar Shinar}, {and} \bibinfo{person}{Craig Schroeder}.} \bibinfo{year}{2024}\natexlab{}.
\newblock \showarticletitle{Higher order divergence-free and curl-free interpolation on MAC grids}.
\newblock \bibinfo{journal}{\emph{J. Comput. Phys.}}  \bibinfo{volume}{503} (\bibinfo{year}{2024}), \bibinfo{pages}{112831}.
\newblock


\bibitem[Saito et~al\mbox{.}(2019)]%
        {saito2019pifu}
\bibfield{author}{\bibinfo{person}{Shunsuke Saito}, \bibinfo{person}{Zeng Huang}, \bibinfo{person}{Ryota Natsume}, \bibinfo{person}{Shigeo Morishima}, \bibinfo{person}{Angjoo Kanazawa}, {and} \bibinfo{person}{Hao Li}.} \bibinfo{year}{2019}\natexlab{}.
\newblock \showarticletitle{{PIFu}: Pixel-aligned implicit function for high-resolution clothed human digitization}. In \bibinfo{booktitle}{\emph{Proceedings of the IEEE/CVF International Conference on Computer Vision}}. \bibinfo{pages}{2304--2314}.
\newblock


\bibitem[Stam(2023)]%
        {stam2023stable}
\bibfield{author}{\bibinfo{person}{Jos Stam}.} \bibinfo{year}{2023}\natexlab{}.
\newblock \showarticletitle{Stable fluids}.
\newblock In \bibinfo{booktitle}{\emph{Seminal Graphics Papers: Pushing the Boundaries, Volume 2}}. \bibinfo{pages}{779--786}.
\newblock


\bibitem[Wang et~al\mbox{.}(2024)]%
        {wang2024physics}
\bibfield{author}{\bibinfo{person}{Yiming Wang}, \bibinfo{person}{Siyu Tang}, {and} \bibinfo{person}{Mengyu Chu}.} \bibinfo{year}{2024}\natexlab{}.
\newblock \showarticletitle{Physics-Informed Learning of Characteristic Trajectories for Smoke Reconstruction}. In \bibinfo{booktitle}{\emph{ACM SIGGRAPH 2024 Conference Papers}}. \bibinfo{pages}{1--11}.
\newblock


\bibitem[Wang et~al\mbox{.}(2004)]%
        {wang2004image}
\bibfield{author}{\bibinfo{person}{Zhou Wang}, \bibinfo{person}{Alan~C Bovik}, \bibinfo{person}{Hamid~R Sheikh}, {and} \bibinfo{person}{Eero~P Simoncelli}.} \bibinfo{year}{2004}\natexlab{}.
\newblock \showarticletitle{Image quality assessment: from error visibility to structural similarity}.
\newblock \bibinfo{journal}{\emph{IEEE transactions on image processing}} \bibinfo{volume}{13}, \bibinfo{number}{4} (\bibinfo{year}{2004}), \bibinfo{pages}{600--612}.
\newblock


\bibitem[Wendland(1995)]%
        {wendland1995piecewise}
\bibfield{author}{\bibinfo{person}{Holger Wendland}.} \bibinfo{year}{1995}\natexlab{}.
\newblock \showarticletitle{Piecewise polynomial, positive definite and compactly supported radial functions of minimal degree}.
\newblock \bibinfo{journal}{\emph{Advances in computational Mathematics}} \bibinfo{volume}{4}, \bibinfo{number}{1} (\bibinfo{year}{1995}), \bibinfo{pages}{389--396}.
\newblock


\bibitem[Wu et~al\mbox{.}(2024)]%
        {wu20244d}
\bibfield{author}{\bibinfo{person}{Guanjun Wu}, \bibinfo{person}{Taoran Yi}, \bibinfo{person}{Jiemin Fang}, \bibinfo{person}{Lingxi Xie}, \bibinfo{person}{Xiaopeng Zhang}, \bibinfo{person}{Wei Wei}, \bibinfo{person}{Wenyu Liu}, \bibinfo{person}{Qi Tian}, {and} \bibinfo{person}{Xinggang Wang}.} \bibinfo{year}{2024}\natexlab{}.
\newblock \showarticletitle{4d gaussian splatting for real-time dynamic scene rendering}. In \bibinfo{booktitle}{\emph{Proceedings of the IEEE/CVF conference on computer vision and pattern recognition}}. \bibinfo{pages}{20310--20320}.
\newblock


\bibitem[Xie et~al\mbox{.}(2025)]%
        {xie2025fluidgs}
\bibfield{author}{\bibinfo{person}{Youchen Xie}, \bibinfo{person}{Chen Li}, \bibinfo{person}{Sheng Qiu}, \bibinfo{person}{Zhi-Jun Wang}, \bibinfo{person}{Chenhui Li}, \bibinfo{person}{Yibo Zhao}, \bibinfo{person}{Zan Gao}, {and} \bibinfo{person}{Changbo Wang}.} \bibinfo{year}{2025}\natexlab{}.
\newblock \showarticletitle{FluidGS: Physics Informed Gaussian Splatting for Dynamic Fluid Reconstruction from Sparse Views}. In \bibinfo{booktitle}{\emph{Proceedings of the 33rd ACM International Conference on Multimedia}}. \bibinfo{pages}{8438--8447}.
\newblock


\bibitem[Xiong et~al\mbox{.}(2017)]%
        {xiong2017rainbow}
\bibfield{author}{\bibinfo{person}{Jinhui Xiong}, \bibinfo{person}{Ramzi Idoughi}, \bibinfo{person}{Andres~A Aguirre-Pablo}, \bibinfo{person}{Abdulrahman~B Aljedaani}, \bibinfo{person}{Xiong Dun}, \bibinfo{person}{Qiang Fu}, \bibinfo{person}{Sigurdur~T Thoroddsen}, {and} \bibinfo{person}{Wolfgang Heidrich}.} \bibinfo{year}{2017}\natexlab{}.
\newblock \showarticletitle{Rainbow particle imaging velocimetry for dense 3D fluid velocity imaging}.
\newblock \bibinfo{journal}{\emph{ACM Transactions on Graphics (TOG)}} \bibinfo{volume}{36}, \bibinfo{number}{4} (\bibinfo{year}{2017}), \bibinfo{pages}{1--14}.
\newblock


\bibitem[Yang et~al\mbox{.}(2024)]%
        {yang2024deformable}
\bibfield{author}{\bibinfo{person}{Ziyi Yang}, \bibinfo{person}{Xinyu Gao}, \bibinfo{person}{Wen Zhou}, \bibinfo{person}{Shaohui Jiao}, \bibinfo{person}{Yuqing Zhang}, {and} \bibinfo{person}{Xiaogang Jin}.} \bibinfo{year}{2024}\natexlab{}.
\newblock \showarticletitle{Deformable 3d gaussians for high-fidelity monocular dynamic scene reconstruction}. In \bibinfo{booktitle}{\emph{Proceedings of the IEEE/CVF conference on computer vision and pattern recognition}}. \bibinfo{pages}{20331--20341}.
\newblock


\bibitem[Yu et~al\mbox{.}(2023)]%
        {yu2023inferring}
\bibfield{author}{\bibinfo{person}{Hong-Xing Yu}, \bibinfo{person}{Yang Zheng}, \bibinfo{person}{Yuan Gao}, \bibinfo{person}{Yitong Deng}, \bibinfo{person}{Bo Zhu}, {and} \bibinfo{person}{Jiajun Wu}.} \bibinfo{year}{2023}\natexlab{}.
\newblock \showarticletitle{Inferring Hybrid Neural Fluid Fields from Videos}. In \bibinfo{booktitle}{\emph{Thirty-seventh Conference on Neural Information Processing Systems}}.
\newblock


\bibitem[Zhang et~al\mbox{.}(2018)]%
        {zhang2018unreasonable}
\bibfield{author}{\bibinfo{person}{Richard Zhang}, \bibinfo{person}{Phillip Isola}, \bibinfo{person}{Alexei~A Efros}, \bibinfo{person}{Eli Shechtman}, {and} \bibinfo{person}{Oliver Wang}.} \bibinfo{year}{2018}\natexlab{}.
\newblock \showarticletitle{The unreasonable effectiveness of deep features as a perceptual metric}. In \bibinfo{booktitle}{\emph{Proceedings of the IEEE conference on computer vision and pattern recognition}}. \bibinfo{pages}{586--595}.
\newblock


\bibitem[Zhu and Bridson(2005)]%
        {zhu2005animating}
\bibfield{author}{\bibinfo{person}{Yongning Zhu} {and} \bibinfo{person}{Robert Bridson}.} \bibinfo{year}{2005}\natexlab{}.
\newblock \showarticletitle{Animating sand as a fluid}.
\newblock \bibinfo{journal}{\emph{ACM Transactions on Graphics (TOG)}} \bibinfo{volume}{24}, \bibinfo{number}{3} (\bibinfo{year}{2005}), \bibinfo{pages}{965--972}.
\newblock


\end{thebibliography}
}



\end{document}